\documentclass[sigconf]{acmart}
\AtBeginDocument{%
  \providecommand\BibTeX{{%
    \normalfont B\kern-0.5em{\scshape i\kern-0.25em b}\kern-0.8em\TeX}}}

\setcopyright{acmcopyright}
\copyrightyear{2024}
\acmYear{2024}
\acmDOI{XXXXXXX.XXXXXXX}

\copyrightyear{2024} 
\acmYear{2024} 
\setcopyright{rightsretained} 
\acmConference[CHI '24]{Proceedings of the CHI Conference on Human Factors in Computing Systems}{May 11--16, 2024}{Honolulu, HI, USA}
\acmBooktitle{Proceedings of the CHI Conference on Human Factors in Computing Systems (CHI '24), May 11--16, 2024, Honolulu, HI, USA}
\acmDOI{10.1145/3613904.3642056}
\acmISBN{979-8-4007-0330-0/24/05}

\usepackage{subcaption}
\usepackage{nowidow}
\usepackage{microtype}

\newcommand{\shortquote}[1]{``#1''}
\newcommand{\worktitle}[1]{\textit{#1}}

\begin{document}

\title{When the Body Became Data: \\ Historical Data Cultures and Anatomical Illustration} 

\author{Michael Correll}
  \email{m.correll@northeastern.edu}
  \orcid{0000-0001-7902-3907}
\authornote{Both authors contributed equally to this research.}
\affiliation{%
  \institution{Northeastern University}
  \city{Portland}
  \state{Maine}
  \country{USA}
  }

\author{Laura A. Garrison}
  \email{laura.garrison@uib.no}
  \orcid{0000-0001-7134-2006}
\authornotemark[1]
\affiliation{%
  \institution{University of Bergen}
  \city{Bergen}
  \country{Norway}
}

\renewcommand{\shortauthors}{Correll and Garrison}

\begin{abstract}
With changing attitudes around knowledge, medicine, art, and technology, the human body has become a source of information and, ultimately, shareable and analyzable \textit{data}. Centuries of illustrations and visualizations of the body occur within particular historical, social, and political contexts. These contexts are enmeshed in different so-called \textit{data cultures}: ways that data, knowledge, and information are conceptualized and collected, structured and shared. In this work, we explore how information about the body was collected as well as the circulation, impact, and persuasive force of the resulting images. We show how mindfulness of data cultural influences remain crucial for today's designers, researchers, and consumers of visualizations. We conclude with a call for the field to reflect on how visualizations are not timeless and contextless mirrors on objective data, but as much a product of our time and place as the visualizations of the past.
\end{abstract}




%

\maketitle

\section{Introduction}
Walking along Brick Lane in London, a visitor may be surprised to stumble upon a larger-than-life mural of a human chest X-ray with a gaping, fractured black cavity in the center where the heart has been removed~\cite{londoncalling2018}. Titled \textit{Stolen Heart}, this piece by street artist SHOK-1 brings anatomical data typically reserved for clinicians in dark rooms into popular culture to drive a social and political commentary via the depiction of the inner body~\cite{SHOK2023}. Quite a few things must be true for this piece to have its intended impact: the anatomical fact of the missing organ must be identifiable, the visual genre of X-ray imaging technology recognized and accepted, and any visceral reaction to seeing inside the body at least partially overcome. And yet, the piece functions as public art, accessible to mass audiences. This mass legibility of internal anatomy is neither universal nor inevitable. How, then, did this process of interpretability happen? How did the body become not just data, but public data? How did the erstwhile taboo and esoteric genre of anatomical visualization become recognizable and legible on a crowded street?

The human body and our interrogation of its internal structures, whether through direct means like observation and dissection or mediated through imaging technology, provide the inputs to the wide-ranging styles and techniques for the visual representation of human anatomy across history. How these anatomical visualizations are acquired, presented, and circulated is founded on particular ideas regarding the sanctity of all, or rather only \textit{some}, human bodies, which serve as a lens for what can and cannot be seen. What kind of information we collect about the body, how we think about this information, and who does the collecting (and from whom) define and constrain the resulting visualizations and illustrations. 

Histories of medical visualization have foregrounded several iconic examples, from Vesalius' muscle-men to the meticulous annotations in \worktitle{Gray's Anatomy}. This focus on individual efforts by lone innovators has a number of negative consequences. For one, it ignores the work of a host of pioneering, hard-working practitioners who have, for historical or sociological reasons, not been placed in this select and illustrious canon~\cite{klein2022data,evergreen2019}. Another issue with this narrow narrative focus is a lack of consideration for how data visualizations are enmeshed and intertwined with cultures of knowledge and evidence, circulation and persuasion. That is, just as \shortquote{the hand-mill gives you society with the feudal lord; the steam-mill, society with the industrial capitalist,}~\cite{marx1920poverty} so does the printing press give you Vesalius' muscle-men, the graphics card interactive volumetric renders. In Aristotelian terms, while the success of a visualization is partially due to the strength of the insights it communicates (the \textit{logos} of the data-driven argument), these insights are only legible, let alone transformative, given situated and serendipitous confluences of factors: the \textit{kairos} of when, where, and to whom this argument appears~\cite{aristotle}. 

In this paper, we review a set of historical visualization and illustration practices in medicine with an eye towards how their production is influenced by changing \textit{data cultures}: our term for the interconnected epistemologies, rhetorics, and technologies associated with information and, eventually, data. Different data cultures impact how information is conceptualized and collected, structured and shared: even the notion of ``data'' is itself a historically recent and contingent term. We focus in this work on how data cultures around anatomy, imaging, and diagnostics have supported the creation or adaption of different visual forms of data representation with differing social impacts. With these changing data cultures, we also explore how and when the human body itself became a source of data. We perform this analysis with the goal of critiquing emerging projects in medical visualization, and visualization more broadly, that are not mindful that \shortquote{our forms of attention, observation, and truth are situated, contingent, and contested and that the ways we are trained, and train ourselves, to observe, document, record, and analyze the world are deeply historical in character}~\cite{halpern2015beautiful}.
We use snapshots of the long and global history of medical illustration to demonstrate how technological, productive, and political forces shape data visualizations. These historical shifts in data cultures occurred in parallel with changes in religious, cultural, and political life, even as the underlying ``data source''---the human body--- kept its shape. The preconditions for these influential data visualizations to arise, and for them to be successful at their rhetorical goals, are the existence of particular sociotechnical milieux: the right epistemologies, technologies, and even sufficiently (bio-)powerful governments. Not only do these milieux provide important preconditions for certain kinds of medical visualizations, but similar forms of medical visualizations, arising in different historical contexts, can have remarkably different impacts. Likewise, these visualizations can function as reactions against prevailing sociopolitical conditions: transitions from personal experiences to impersonal datasets (and vice versa), from experts to mass audiences, constrained by (or intentionally and iconoclastically flouting) convention and taboo. 

Prior works have explored the history of visualization development with a broad perspective, seating the earliest germinations of visualization in cartography and developing through the centuries into visual depictions of statistics and other scientific disciplines in step with advancements in technology, mathematics, and observational practice~\cite{Friendly06hbook,friendly2021history,marchese2012origins,marchese2013medieval}. Projects and compendia such as the \worktitle{Milestones Project}~\cite{friendly2001milestones} and \worktitle{History of Information Graphics}~\cite{rendgen2019historyinfographic} explore key figures and exemplars of visualization through history, again spanning numerous disciplines. Contemporary exhibitions such as the University of Toronto's \worktitle{Emerging Patterns: Data Visualization Throughout History}~\cite{ut2023emergingpatterns}, show a keen public interest in historic visualization, their patterns of rhetoric, and the role of technology in visualization practices. In contrast to these broad-ranging works that seat the foundations of visualization in maps of land and stars, our analysis focuses on the maps we make of the human body and our health. While others have explored the changing maps of the body, they observe these changes through the lens of artistic or medical practice~\cite{sappol2004morbid, sappol2006dream,reichle2009art,ghosh2015evolution,ruiz2016tedmed}, our lens shifts to view the connection between the resulting image and the data culture that led to this image. Rather than presenting a standalone set of pioneers or ``firsts,'' we also focus on the data cultures surrounding our examples instead of taking a final graphic as a given, or as part of a teleological process of increasing sophistication or forward progress through time.

\subsection{Organization \& Methodology}

The remainder of the paper is organized into a series of historical \textbf{vignettes} about how we have visualized the human body as an object of study and, eventually, as data. What we mean by a ``vignette'' is a thematically organized collection of visual and narrative examples, with associated commentary, meant to highlight a particular theme. The vignette structure allows us to foreground our critiques without constraining ourselves to an exhaustive historical overview. We situate this approach within other commentaries on the historical and political changes of technological concepts that have similar structures: for instance, Wysocki \& Johnson-Eilola's~\cite{wysocki_blinded_1999} use of ``bundles'' of thematically linked examples and commentary on the subject of literacy as a universalizing metaphor for technical aptitude, or Walter Benjamin's~\cite{benjamin1999arcades} exploration of the sociopolitical implications of Parisian architecture through interconnected prose, poetry, and commentary in his \worktitle{Arcades Project}.

Each of our vignettes contains a collection of stories and representative imagery aligning along a central theme that ties human perspectives to technological and societal norms in the visual representation of the human body and disease. Each vignette is built around the same basic structure. We begin with some historical context of the vignette's central theme: the spiritual, philosophical, moralistic underpinnings, state of technology, or productive forces at play. We then introduce a set of representative visual examples, and conclude with a discussion of the applicability, resonance, or remaining dangers of this theme as applied to contemporary visualization practice.

Our choice of historical examples was anchored on our previously mentioned set of ubiquitous examples from medical history (e.g., Galen, Vesalius, and Gray). In service of our goal of highlighting lesser-known examples, we expanded this corpus by consulting public archives such as the U.S. National Library of Medicine's \worktitle{Historical Anatomies on the Web} collection as well as the Wellcome Collections' digital library of historical works related to health and human experience. Our contemporary examples were sourced from core visualization and HCI venues. This search was combined with field research in the context of visiting museums with medical history and visualization-focused exhibitions to collect relevant works and key figures in this space, and finally augmented with examples encountered in our sourcing and literature review of prior works around visual cultures and anatomical history.

\begin{figure*}
  \centering
 \includegraphics[width=\linewidth]{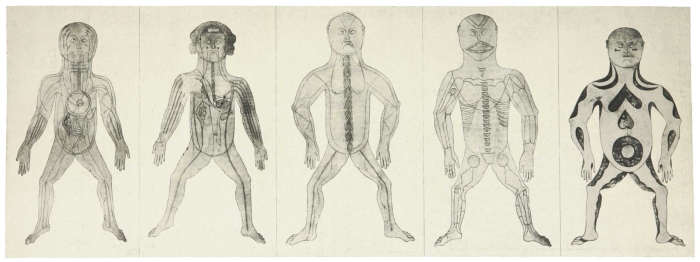}
  \caption{\worktitle{Five Figure Series}, one of earliest known anatomical diagrams, the original of which possibly dating to 300 B.C.~when human dissection was permitted in Alexandria. Each panel describes one of five anatomical systems (as described by Galen): arteries, veins, nerves, bones, and muscles~\cite{standring2016brief}. The human form is arranged as if prepared on a dissection table, in an open, partially-squatting pose.}
  \Description{Five juxtaposed human figures each with interior lines and shapes schematically representing one of five anatomical systems.}
 \label{fig:fivefigures}
\end{figure*}

We present our vignettes across both temporal and cultural shifts in data cultures. The first vignette deals with early attempts to visualize the body in the absence of information from human dissection: that is, under conditions of data scarcity. The second vignette deals with how philosophies and technologies of seeing influenced the resulting illustrations of bodies even once this information was more widely available. The third vignette explores how conceptions of disease likewise resulted in new genres of anatomical illustration. The fourth deals with the popularization and circulation of these images to mass audiences. Finally, the last vignette deals with the ethical and sociopolitical ramifications of the process of reducing the human body down to data.

Taken together, we use these vignettes to argue for a more expansive conception not just of medical or anatomical visualizations, but of visualizations generally. There are no ``neutral'' choices when it comes to representing the body: the process of turning bodies into data to be visualized is inherently subjective and political. We also call on the visualization community to revisit the assumption that the history of visualization is an inherent progression from the simple to the complex, the less accurate to the more accurate, but to consider how our forebears dealt with challenges of data access and rhetoric, complexity and uncertainty. We find that many of the visual forms and genres of the past reappear, clothed in new bodies, to solve perpetually recurring problems in visualization and information design.

The study of medical visualization can and has occupied entire books and careers, and we do not intend for this paper to provide a complete or comprehensive overview of the subject, even within the comparatively narrower lens of cultures of data and their impact on visualization practices in anatomy. For one, knowledge is situated and limited by perspectives. While we have sought out sources beyond the usual---often Western-focused---history of medicine, our coverage of these sources is comparatively limited. For another, we as co-authors are situated within HCI and computer science contexts, and so our analytical focus on our examples as technological artifacts or displays of information provides less historical context than in other treatments of these same examples. Lastly, while many of the examples in our vignettes reward deep contemplation or analysis, for reasons of space we devote relatively little time to each example to instead focus on how these examples are situated within the context of our larger arguments. We invite the reader to take time to assess the figures (or examples from their own practice) in greater depth than is afforded in this work.

\section{Inferring Human Anatomy}
\label{sec:inferring}

In this first vignette, we explore how societies went from \textit{inferring} human anatomy to physically \textit{engaging} with human anatomy. That is, how visual approaches to the body arose under conditions of what we term ``data scarcity'': limited systematized ways of structuring, collecting, or disseminating anatomical information to medical audiences. We begin with a brief overview of the attitudes and perspectives that shaped anatomical and medical visualizations of antiquity. Many early societies created rich and intricate anatomical diagrams based on sources of information or epistemological structures that are different from the structures or notions of data that we associate with modern scientific empiricism. 
Many of these early choices in the systematization and representation of anatomy remain relevant today.  

\begin{figure}
    \centering
   \begin{subfigure}{0.75\linewidth}
         \centering
         \includegraphics[width=\textwidth]{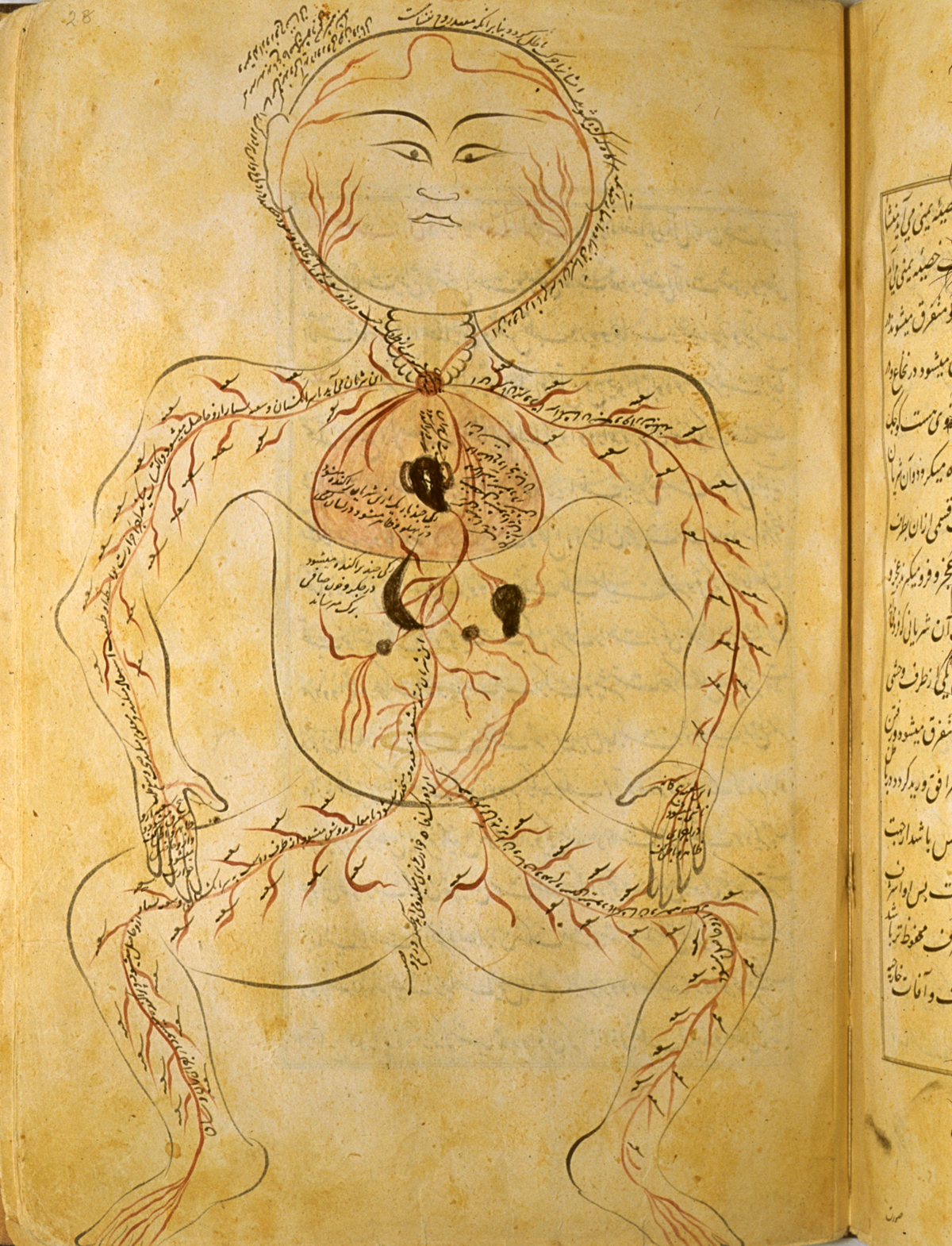}
         \phantomsubcaption
          \Description{Diagram of a squatting male figure with arterial structures shown in opaque, red watercolor.}
         \label{fig:mansur_male}
     \end{subfigure}
     ~\\
        \begin{subfigure}{0.75\linewidth}
         \centering
         \includegraphics[width=\textwidth]{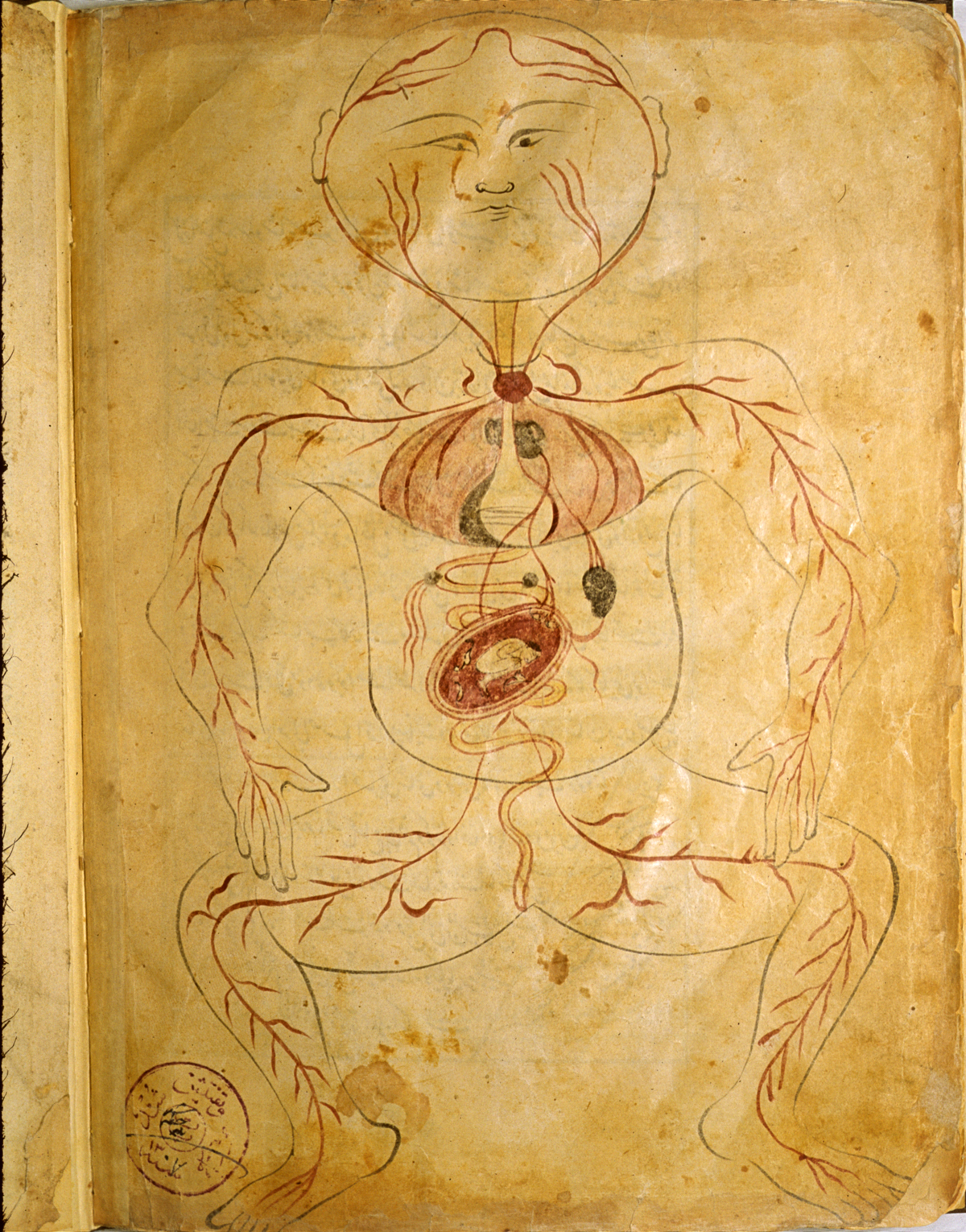}
         \phantomsubcaption
          \Description{Diagram of a squatting female figure with arterial structures shown in opaque, red watercolor and a gravid uterus with the fetus in a breech (transverse) position.}
         \label{fig:mansur_female}
     \end{subfigure}
    \caption{The earliest known color anatomical illustrations of human body from the Islamic world, produced by Man\d{s}\={u}r ibn Ily\={a}s in 1390. He organizes the body according the Greek five systems (per Figure~\ref{fig:fivefigures}), with the arterial system depicted in male (top panel) 
    and female (bottom panel) 
    anatomy. The female illustration is identical to the male, apart from the inclusion of the uterus with a fetus in breech position. The labels for the male appear in a mixture of Arabic and Persian, yet are absent in the female version. This work is also one of the earliest surviving examples of the use of color to strategically encode different anatomical structures and systems.
    }
    \label{fig:mansur}
\end{figure}

Human dissection has been rife with taboos and ethical questions throughout history that provide the limits for what can and cannot be visualized through direct observation~\cite{standring2016brief}. Early societies viewed the human body through symbolic, social and religious lenses. These views were encoded into laws and customs that precluded human cadaver dissection for centuries~\cite{von1992discovery,matuk2006seeing,akkin2014glimpse,luesink2017anatomy}, with a few rare and notable exceptions to this moratorium in the Babylonian and Persian empires~\cite{shoja2007history} and during the Ptolemaic reign in Alexandria~\cite{ghosh2015human,von1975experiment,standring2016brief}. These brief windows provided physicians a direct window to the body's workings, although preservation technology of the time imposed limits on the scope and nuance of dissection. Visual reproductions of the perception of the interior body from this period survive in the \worktitle{Five Figure Series} (Fig.~\ref{fig:fivefigures}), one of the earliest known illustrations of the human body. Here, the body is arranged in a so-called frog pose, with the human form arranged as if on a dissection table in an open, squatting posture. Each panel depicts one of the five perceived bodily systems: arterial and venous blood, nerves, bones, and muscles, respectively~\cite{standring2016brief}. 

Through networks of trade, manual reproductions and variations of these symbolic figures diffused into other regions. Reaching the Islamic world, the illustrations of fourteenth-century C.E.~Man\d{s}\={u}r ibn Ily\={a}s, the earliest known illustrations of the human body from this region, are thought to have in part been inspired by the \worktitle{Five Figure Series}~\cite{shoja2007history}. Beautifully colored with annotated structures on several plates, e.g., Fig.~\ref{fig:mansur}, Man\d{s}\={u}r's work is also thought to blend local religious culture with dissection traditions espoused by Galen. For over a thousand years, Galen, the overwhelmingly prolific second-century C.E.~anatomist and physician, was at the center of the largely unassailable canon of Western medicine, alongside the imposing figures of Greek philosophy and medicine Aristotle and Hippocrates~\cite{zimmer2005soul,standring2016brief}. 
 
\begin{figure}
  \centering
 \includegraphics[width=\linewidth]{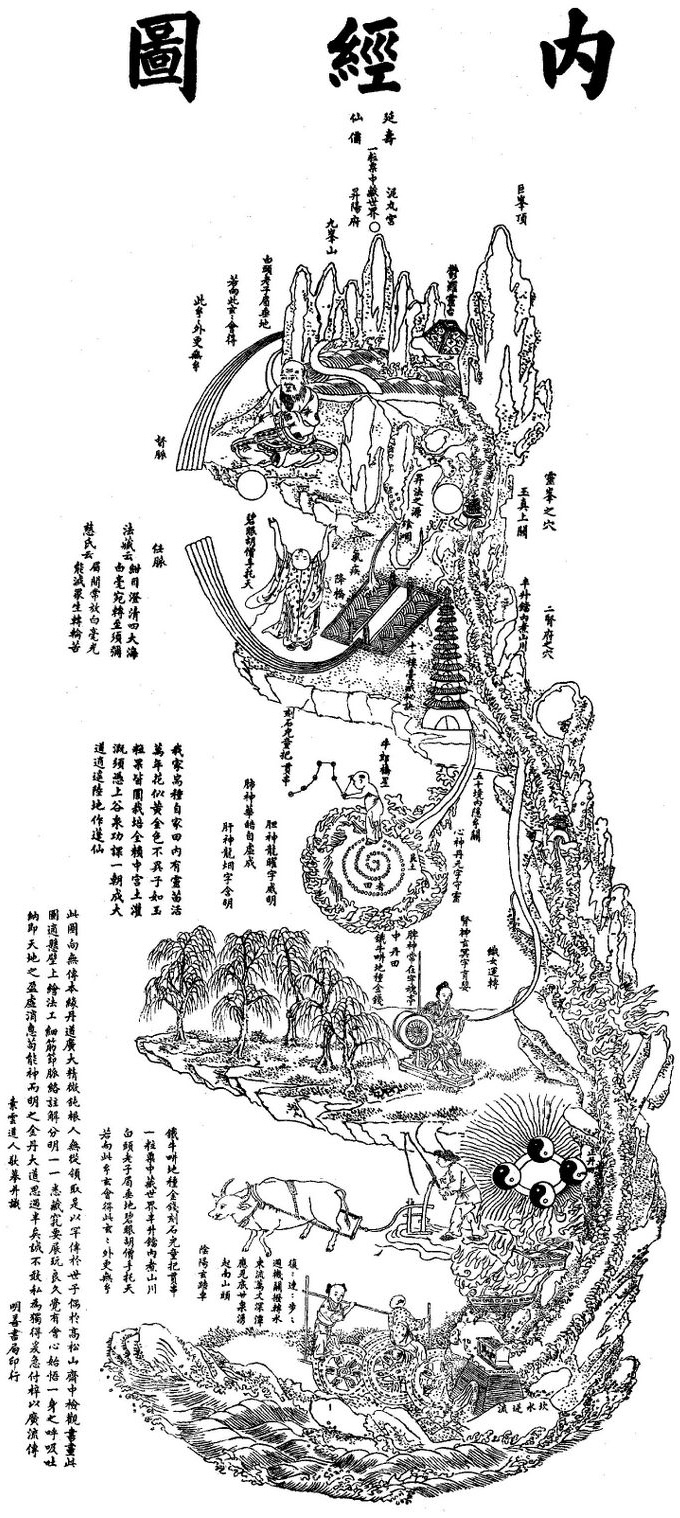}
  \caption{\worktitle{Nei Jing Tu} (Chart of the Inner Landscape) from 1886 that builds on Taoist philosophies of the human body as a microcosm of the universe, using the rough outline fetal form for structure. The head contains a glade in the heavenly Kunlun mountains where direct communication with the divine is possible, with other portions corresponding to both earthly and cosmological sites.~\cite{ramellini2013life}\noindent
  }
  \Description{Illustration of the inner realm as forest in series of levels with different spiritual symbols throughout the image.}
  \label{fig:neijingtu}
\end{figure}

The focus of these Greek and Islamic anatomical charts and similar early diagrams, driven by both the conditions of data scarcity but also their associated philosophical underpinnings, was on the major connections within bodily systems and their contribution to a single whole, rather than precise anatomical structures or locations. These schematic diagrams echo the conventions and distortions of the modern-day subway map, where connections between important components are of greater importance than specific location. A similar influence of epistemology on form can be seen in Eastern anatomical diagrams as well. Zoroastrian, Taoist (Fig.~\ref{fig:neijingtu}), and Buddhist traditions saw bodies as complete inner worlds with their own metaphorical forests, stars, and beings~\cite{matuk2006seeing,shoja2007history}. As with our prior examples, the visual emphasis is again on a cohesive whole, but the physical reality of the subject is abstracted away to instead focus on the metaphorical structure of the body. 

Other artifacts in the visual representation of bodies were due to the use of animals as substitute sources of information, with the assumption that the structures found in other animals when they were dissected could or would be similarly represented in humans~\cite{von1992discovery,zimmer2005soul}. Galen, who in his lifetime dissected many animals but likely never humans (in part due to the aforementioned taboos on human dissection), produced occasional inaccuracies that were noted by anatomists centuries after the fact. Per Singer~\cite{singer1949galen}, \shortquote{Roughly speaking, we may say his anatomy is that of the soft parts of the Barbary ape, \textit{Macaca Inuus}, imposed on the human skeleton.} Misconceptions about the human body (such as the diagnosis of ``hysteria'' as a result of a literal ``wandering womb'', i.e., a displaced uterus~\cite{faraone2011magical}), in the absence of newer or more authoritative data, therefore lingered for centuries in the Islamic and Christian worlds~\cite{standring2016brief}. For instance, Galen's claims that the liver spread out like a hand led to anatomists mistakenly assuming that the human liver had five lobes into the 16th century~\cite{mcclusky1997hepatic}. Some of these misconceptions based on animal dissections remained in later anatomical texts even after human dissection was more commonplace~\cite{singer1946some}. 

\begin{figure}
    \centering
   \begin{subfigure}[b]{0.45\linewidth}
         \centering
         \includegraphics[width=\textwidth]{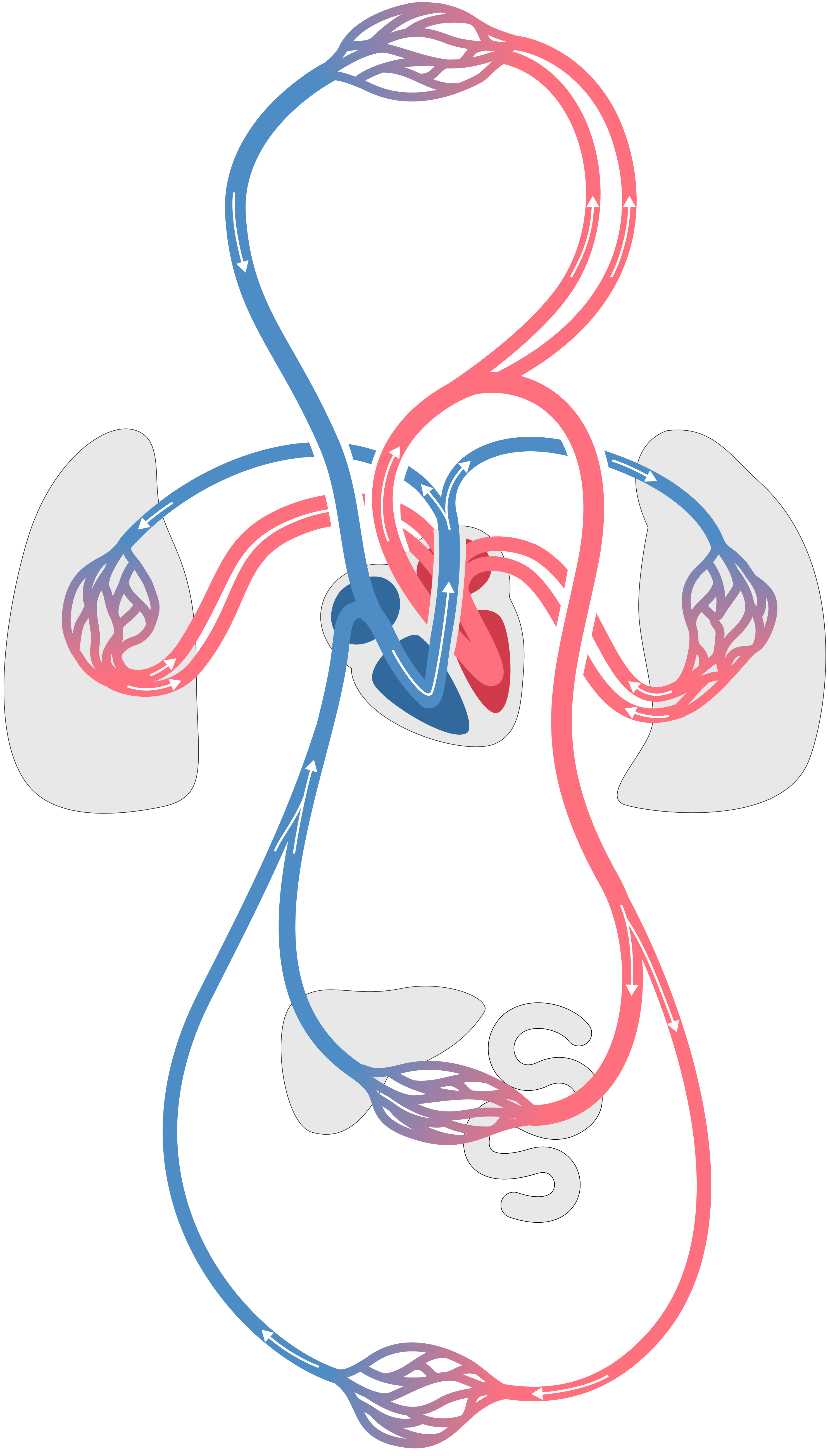}
         \phantomsubcaption
          \Description{Diagrammatic anatomical illustration of the circulatory system. A cross section of the heart resides in the middle of the diagram, with paired vessels (red for arteries, blue for veins) extending in each of the four cardinal directions from the heart. The top pair join in one capillary loop to show oxygen exchange in the head, while the bottom pair similarly join in two places to show oxygen exchange in the internal organs and legs, respectively. The left and right vessel pairs extend to eventually join in the left and right lungs, respectively.}
         \label{fig:circulatorydiagram}
     \end{subfigure}
     \hfill
     \begin{subfigure}[b]{0.42\linewidth}
       \centering
       \includegraphics[width=\textwidth]{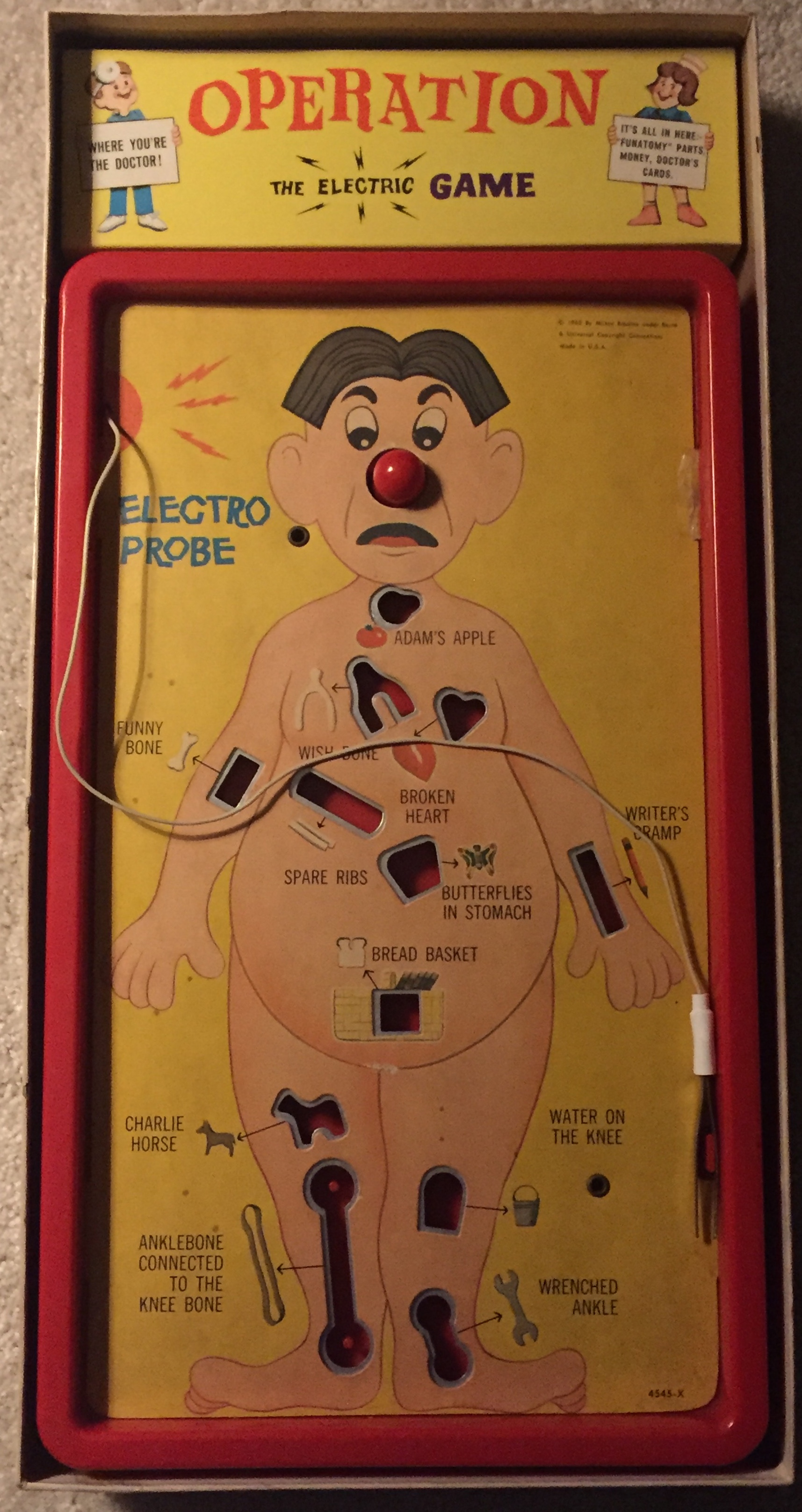}
       \phantomsubcaption
       \Description{Photograph of 1965 children's game, Operation, showing a cartoon male figure with a large red clown nose and cutouts over different regions of his body for the placement of of satirical body parts, e.g., a funny bone.}
       \label{fig:operation}
     \end{subfigure}
    \caption{Two examples of modern schematized and simplified representations of anatomy.
    \textit{Left panel:} Schematic illustration of the human circulatory system, styled after what is commonly seen in anatomical textbooks such as \worktitle{Netter's Atlas of Human Anatomy}~\cite{netter2022netter}. Rather than depicting the cardiovascular system precisely as it appears in the body, this work intentionally abstracts away information that does not support the author's didactic aim to describe to the viewer the circular nature of blood flow and oxygen exchange in the human body.
    \textit{Right panel:} \worktitle{Operation} children's game (1965). Designed for children to entertain and engage with anatomy and medicine, the visuals of this game are heavily abstracted from reality in an effort by the game designers to reduce the amount of information down to what is salient and relevant for the target audience. Graphical symbols rather than the true-to-form anatomy are more pertinent to the objective of the visual, similar to the stylized, symbolic anatomical visuals of antiquity.
  }
    \label{fig:modernschema}
\end{figure}

The design choices made under conditions of data scarcity remain relevant today, even in the face of the opposite problem: one of information overload. For instance, a modern-day student of anatomy or medicine has myriad data sources to choose from to understand the structure of the circulatory system: cadaveric dissection, magnetic resonance imaging (MRI), X-ray imaging, etc. Lacking sufficient prior knowledge for what to look for, they may quickly become overwhelmed and fail in their study objective. Schematic, stylized illustrations serve as maps to orient the viewer to the jumble of complex anatomy, abstracting away information deemed irrelevant for the task to focus instead on the essentials. Many anatomical atlases, e.g., \worktitle{Netter's Atlas of Human Anatomy}~\cite{netter2022netter}, feature schematic plates that navigate this information abundance and distill the content down to its essence, as in the circulatory system diagram depicted in Fig.~\ref{fig:modernschema} (left). Similar schematic, iconographic anatomical drawings surface in other contexts, e.g., the children's game \worktitle{Operation}, providing a playful and simplified means to explore and conduct surgery on the body (\autoref{fig:modernschema}, right). 

Finally, we should note that the concept of data scarcity is not one limited to the historical past, and that societies do not inevitably or uniformly move from less data to more data. In spite of a modern-day information abundance, there remain medical phenomena that escape visualization with the data available. A migraine aura is difficult or impossible to capture via imaging data in real-time, due to the unpredictable nature of their onset and an incomplete understanding of their exact pathophysiology~\cite{kikkeri2022migraine}. Instead, clinicians and researchers rely on patients relaying their experience of the (often) visual disturbance of an aura that manifests in, e.g., spots, flashes, zig-zags, among other possible sensory symptoms. Absent acquired data, these qualitative interviews are translated by visualization designers who infer gaps in the data to produce highly conceptual illustrations and photo-manipulated collages that fill in the information necessary to communicate the experience of a migraine aura. Thinking more broadly on absent data in modernity, Onuoha's~\cite{Collier2022Review,onuoha2016library} art installations around the theme of a \worktitle{Library of Missing Datasets} highlight examples where inequalities in race, wealth, and power leave us with very little information about matters of key importance. For instance, there is limited race-based medical information available for pregnancy risks. 

\section{Pedagogies of Sight}
\label{sec:pedagogies}

\begin{figure}
    \centering
   \begin{subfigure}[b]{0.7\linewidth}
         \centering
         \includegraphics[width=\textwidth]{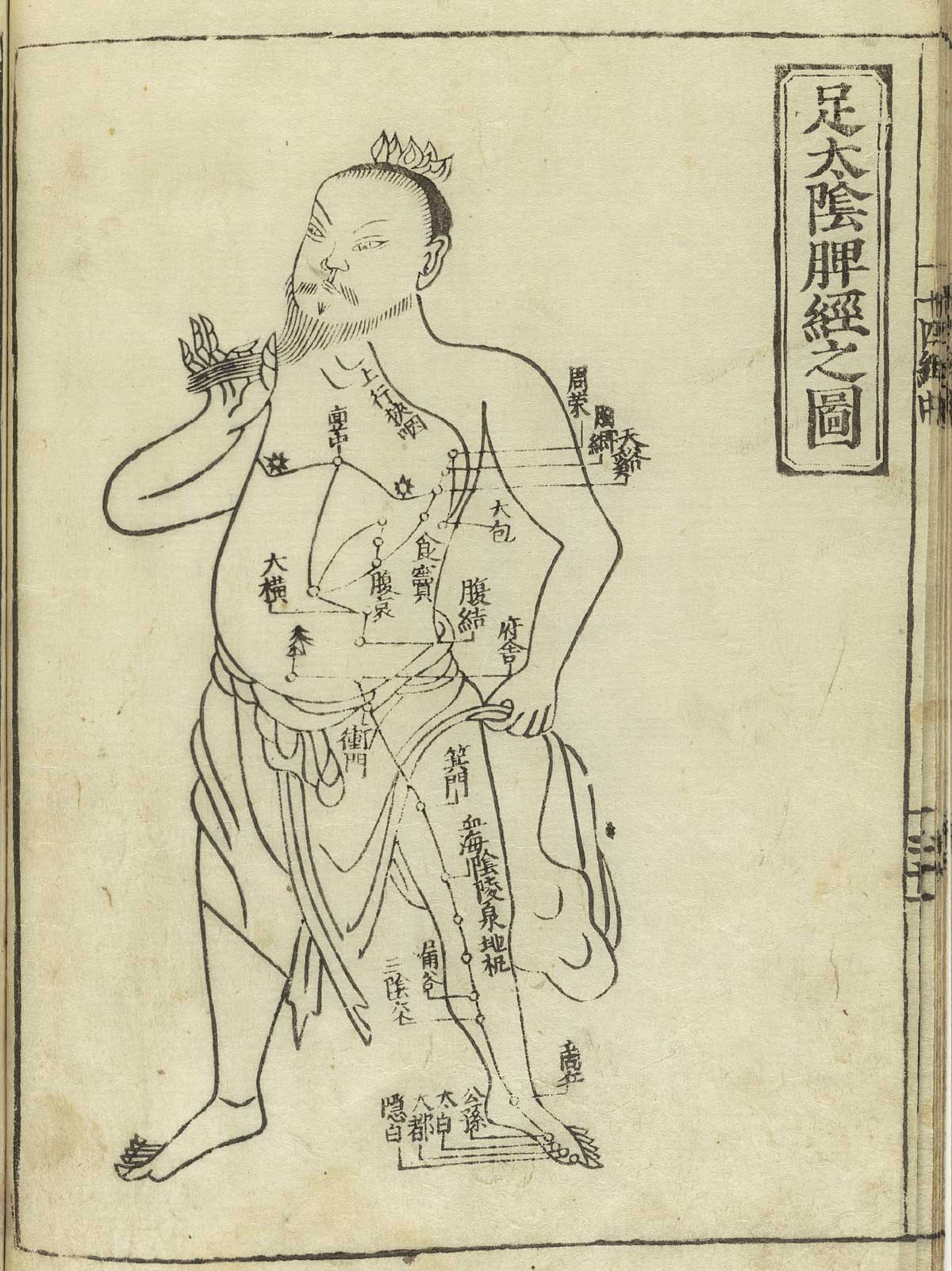}
         \phantomsubcaption
          \Description{Drawing of a man with labeled lines representing the flow and meridians of qi throughout the body.}
         \label{fig:acupuncture}
     \end{subfigure}
     \hfill
     \\
     \begin{subfigure}[b]{0.7\linewidth}
         \centering
         \includegraphics[width=\textwidth]{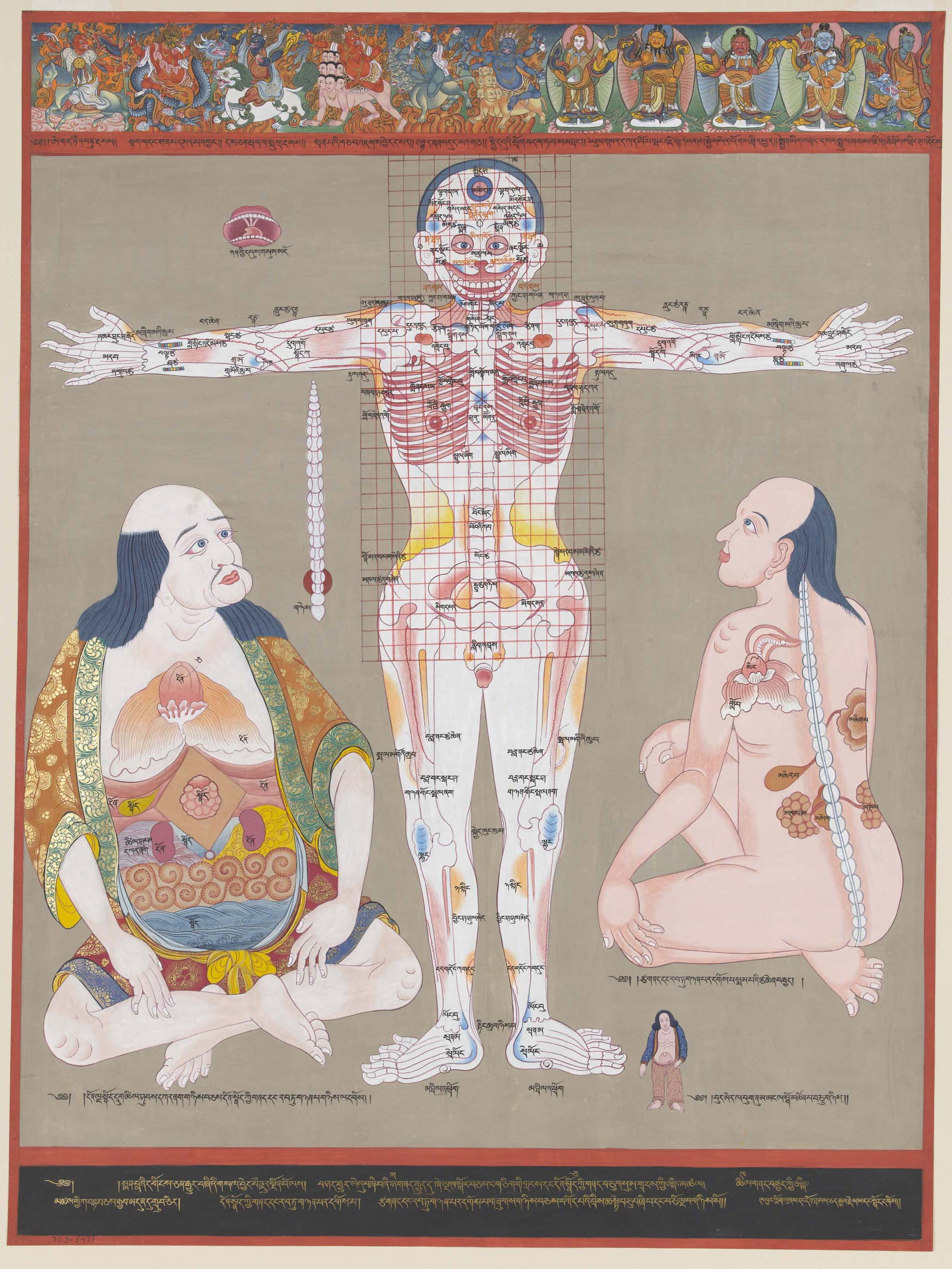}
         \phantomsubcaption
          \Description{Colorful drawing of a human figure showing various important veins and arteries in the body. Figures kneeling in front and behind the figure show major organs.}
         \label{fig:tibetanatomy}
     \end{subfigure}
    \caption{Two anatomical illustrations from China and Tibet, respectively. 
    \textit{Top panel:} Illustration used for acupuncture practice from \worktitle{Routes of the Fourteen Meridians and their Functions} by Hua Shou, 14th century Chinese physician. The focus on lines of flow mean that other anatomic structures, such as musculoskeletal anatomy, are not rendered). 
    \textit{Bottom panel:} 17th-century \textit{thangka} showing annotated vulnerable anatomical points for illness and injury. The standing figure is annotated with the vulnerable blood vessels, muscles, and bones, while the kneeling figures show vulnerable internal organs from the front (left figure) and back (right figure). Again, the focus on vulnerable points de-emphasizes other anatomical features like muscles or bones. 
    }
    \label{fig:easternanatomy}
\end{figure}
Part of the process of both designing and interpreting visualizations is learning \textit{how to see}: choosing which structures to highlight, what extraneous or irrelevant detail to remove, and matching the expected visual genres and conventions of one's audience. Different ways of seeing can produce radically different representations of the same bodily forms. This process of learning to see occurs not just at the level of the individual student: new technologies create new affordances and new views and experiences of the body that need to be rhetorically and socially integrated into the prevailing data cultures. In this vignette, we explore how different data cultures produce different views of the same structures, and how new technologies or relationships to data result in new rhetorical needs to convince audiences of the utility and authority of new ways of seeing. We argue that the depiction of bodies is not a universal and objective process, but a long-running sociotechnical dialogue. 

Taoist concepts of the body include the balance and flow of \textit{yin} and \textit{yang}, and the course of \textit{qi} (the body's vital energy) throughout~\cite{matuk2006seeing, shoja2007history}. This holistic ideal of the body corresponded to visualizations steeped in spiritual metaphor to convey the complex inner world of the body (Figure~\ref{fig:neijingtu}). Other visualizations were created for practical clinical use, such as diagrammatic illustrations of pulse and acupuncture points, or the winding paths of \textit{qi} (Figure~\ref{fig:easternanatomy}, top panel). These works are echoed in the restraint and consideration a modern designer uses in terms of what information to show. Illustrating all acupuncture points would saturate the visual, so the designer must be selective and show only information that supports their rhetorical aims. Musculoskeletal morphology is absent from these visuals---this anatomy was not an important part of the Chinese medical canon and considered irrelevant for medical interventions like acupuncture: indeed, for centuries, physicians lacked words to describe the muscles of the body~\cite{nlmhua}. In the West, a corresponding lack of focus on certain systems and features---in this case on circulatory rhythms---led to a paucity of words to describe heartbeats, with even Galen forced to use highly subjective terms like \shortquote{mouselike} or \shortquote{ant-crawling} to describe irregular pulses~\cite{kuriyama1999expressiveness}. In later centuries, Chinese illustrations remained largely diagrammatic in a reflection of their didactic priorities. Specific organ or muscle morphology was usually irrelevant to the illustrator's overall goals, and so was vaguely shown, if at all, as a way to manage information complexity and to clarify the learning objective~\cite{van1995oriental}. 
Tibetan culture similarly relied on holistic views of the body, often connecting medicine and theology~\cite{sabernig2017vulnerable}. 17th-century Tibetan Buddhist paintings, called \textit{thangkas}, sometimes incorporated medical knowledge from India, ancient Greece, Persia, pre-Buddhist Tibet, and China. Anatomical representations integrate symbolic representations of anatomy with Tibetan life and religion through series of small figures arranged in rows in all or part of the paintings (\autoref{fig:easternanatomy}, bottom panel, and \autoref{fig:tibetandiagnostics}), emphasizing both the spiritual and physical components of the human body~\cite{williamson2009body}. 

While Western medical philosophies began from similar traditions as in the East, including similar metaphors of the body as microcosm of the world~\cite{matuk2006seeing}, and similar medical practices based on the flow of vital energies~\cite{singer1949galen}, divergences in the specifics of these philosophical underpinnings emphasized different parts of the body and so resulted in differing diagrams~\cite{kuriyama1999expressiveness}. For instance, early Greek society valued strength and vitality as physical traits of the divine, and so placed specific focus on musculoskeletal anatomy~\cite{matuk2006seeing}. Some artists rising from later generations enmeshed in this culture went so far as to negotiate commissions with payment in human cadavers~\cite{ebenstein2016anatomical}. Leonardo da Vinci captured the value many artists placed on direct muscular observation and interaction: 
\begin{quote}
    \textit{A good painter must know what muscles swell for any given action, and must emphasize the bulging of those muscles only and not the rest, as some painters do who think that they are showing off their skill when they draw nudes that are knotty and graceless–mere sacks of nuts}~\cite{standring2016brief}.
\end{quote}

\begin{figure}
  \centering
 \includegraphics[width=\linewidth]{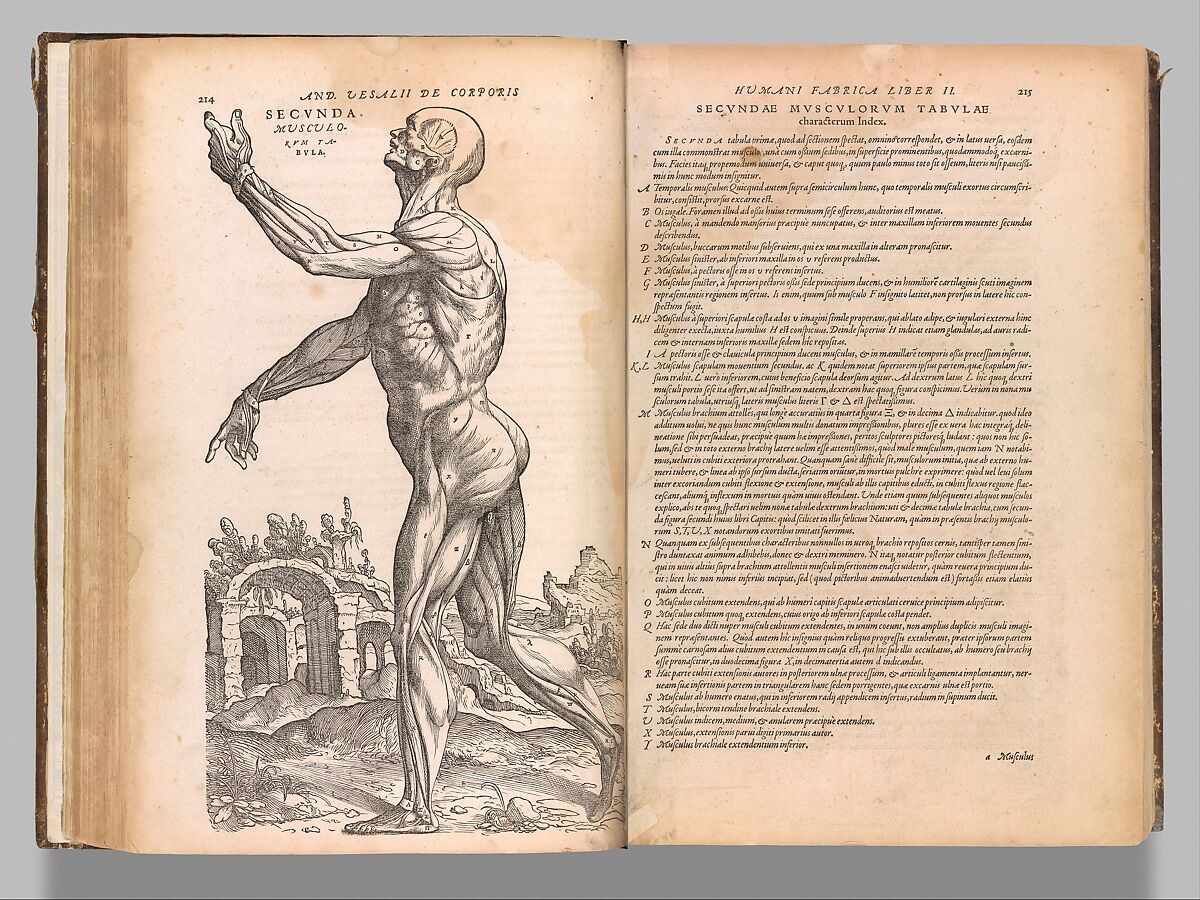}
  \caption{The muscle-men woodcut illustrations from Vesalius' \worktitle{De Humani Corporis Fabrica}, published in 1543, are highly-detailed anatomical illustrations resulting from collaboration between anatomists and artists combined with advancements in printing technology. The legend fills the right facing page. Note the clearly defined muscular structures; other illustrations from the same volume show the muscular figures holding their own skin in their hands while making similar dynamic poses, eyes skyward.}
   \Description{A drawing of a man with skin removed, exposing his clearly-defined muscular system. He gestures with one hand and points with the other.}
  \label{fig:musclemen}
\end{figure}

The seismic political, social, and religious upheavals of the Renaissance brought a culture of curiosity and observation that led anatomists from their distant chairs and tomes to the dissection table~\cite{standring2016brief}. These anatomists began dissecting cadavers themselves, previously unheard of, and describing what they saw. Illustrations of anatomy came to be instructional in their own right, with increased realism thanks to the development of linear perspective as a new way of representing three dimensions on a two-dimensional plane ~\cite{zimmer2005soul}. Some visualizations were even interactive: bone- or wax figures from the tradition of religious reliquaries and effigies had removable internal parts for students to learn by disassembly and reassembly (Figure \ref{fig:venuses}, right panel). ``Fugitive sheets'', anatomical illustrations with paper flaps that could be lifted to reveal underlying anatomy, became wildly popular study aids. The most impressive and detailed of these were distributed by Andreas Vesalius, the Flemish surgeon and anatomist behind the landmark \worktitle{De Humani Corporis Fabrica}, a comprehensive atlas of the human body~\cite{ebenstein2016anatomical} that saw text accompanied by full-page didactic anatomical illustrations for the first time in modern print culture~\cite{xiang2023role} (Figure~\ref{fig:musclemen}). The year of its publication (1543) saw massive paradigm shifts from received wisdom to logic and empirical reasoning on several scientific fronts---Copernicus' heliocentric theory of the solar system was published the same year. However, even Vesalius' work was enmeshed in existing epistemic structures. For instance, the humoural theory of medicine championed by Galen and other forebears requires that organs both emit and absorb humoural fluids, and so Vesalius points to the pituitary gland as an emitter of phlegm, perforations in the interventricular septum of the heart, the existence of the rete mirabile (from Latin: the \textit{wonderful net}) at the base of the brain, even though direct observation did not support such structures or functions in the human body~\cite{singer1949galen,lanska2015evolution}.

The advent of new imaging methods for human anatomical data necessitated new approaches to seeing, sharing, and promoting the validity of these methods. Separated by two centuries, Robert Hooke and Santiago Ram\'{o}n y Cajal created their own ``pedagogies of sight''~\cite{jack2009pedagogy} to help themselves, their scientific colleagues, and the broader public to access the invisible worlds revealed by microscopy and to generate support for this new science~\cite{fiorentini2011inducing}. The first illustrations and visualizations of microscopy images had to perform the dual jobs of not just presenting representations of what was seen, but also of convincing the audiences that the new techniques produced novel but valid insights anchored in past understanding.

This process of teaching, publicizing, and convincing also occurred with more recent imaging techniques like the X-ray and MRI. The development of these imaging techniques flipped traditional means of seeing the body, enabling an \shortquote{inside-out view as opposed to outside-in perspective}~\cite{standring2016brief}. Often, weird is what sells: Wilhelm R\"{o}ntgen, the accidental inventor of the X-ray, famously sent the first X-ray image of his wife's shadowy, bony hand as the family's Christmas card in 1895, similar to how today's \textit{\#AcademicTwitter} users broadcast short gifs and stunning renderings of new imaging methods for thousands to admire (and subsequently queue up for data access). This marketing maneuver generated a hype and adoption frenzy such that, only five years later, the X-ray was firmly planted in clinical diagnostic and treatment practice~\cite{howell2016early}. These technologies ultimately gave rise to the radiology profession, with individuals trained to see and interpret the anatomy of the living. Of note is that this initial rhetorical work may have functioned \textit{too} well: per Daston \& Galiston~\cite{daston1992image} \shortquote{the very form of X-ray photography was a threat because the photographic medium fairly radiated authority, even while practitioners of the art frequently confronted its deceptiveness.} Indeed, contemporaneous sources warned against the potential of the false objectivity of this new way of seeing~\cite{white1900report}:

\begin{quote}
\textit{The routine employment of the X-ray in cases of fracture is not at present of sufficient definite advantage to justify the teaching that it should be used in every case. [...] There is evidence that in competent hands plates may be made that will fail to reveal the presence of existing fractures or will appear to show a fracture that does not exist.}
\end{quote}
This alternation between representational and rhetorical, diagrammatic and photographic, continues in modern science imaging~\cite{richards2003argument}, where photographic images are juxtaposed with diagrams meant to reinforce (visual) arguments and convey scientific authority, and in public-facing comics and cartoons meant to communicate hygienic guidelines as well as the impacts of public health crises like the COVID-19 pandemic~\cite{kearns2020role}. 

\begin{figure}
    \centering
    \includegraphics[width=0.8\linewidth]{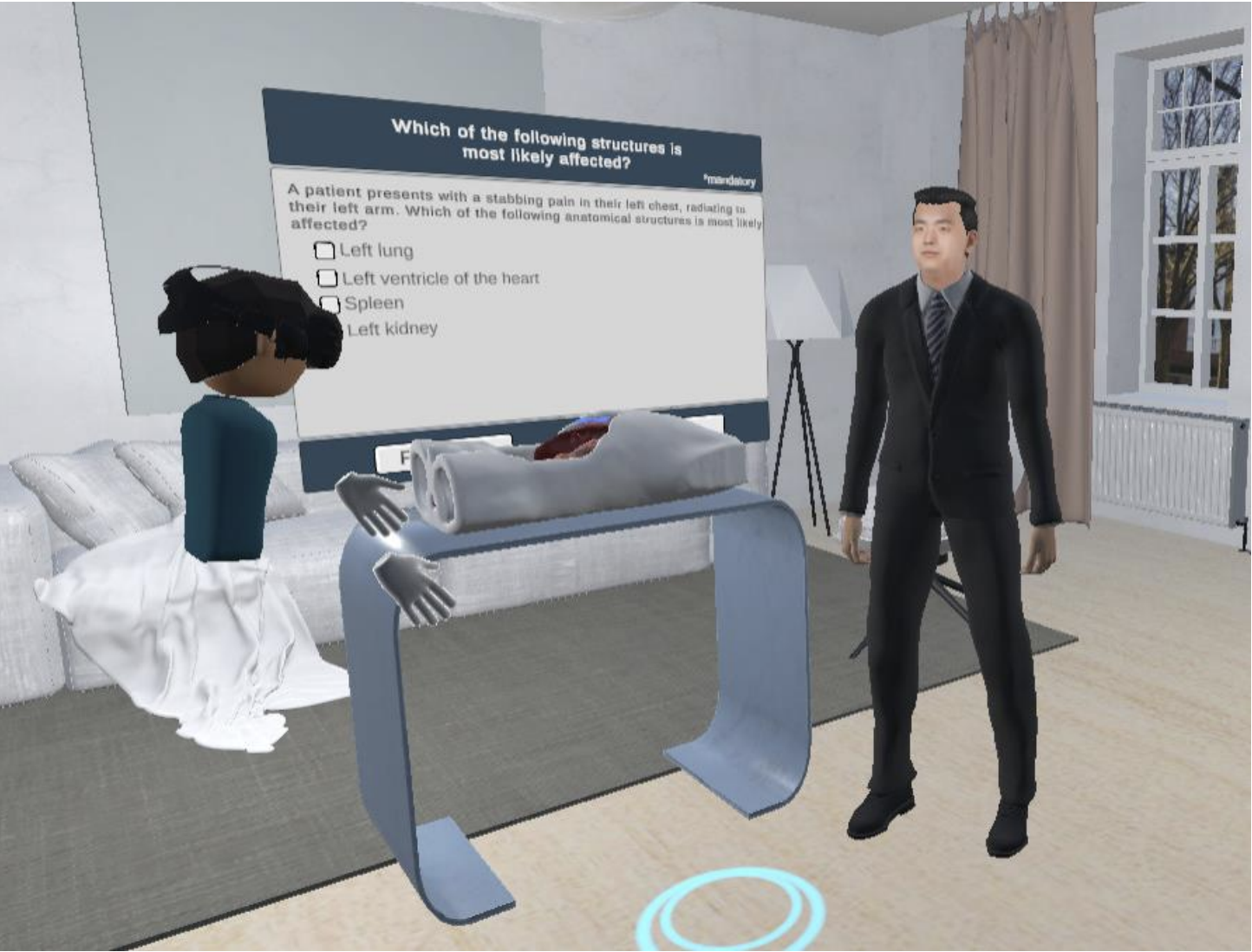}
    \caption{Interface screenshot of an immersive VR environment for human anatomical education, including an embodied photorealistic human virtual assistant integrated with the generative AI service ChatGPT~\protect\cite{chheang2023towards}. The authors' work for this visualization is two-fold: in the immersive VR environment, they must show the user of the ease of interaction and pedagogical parity with the physical dissection environment. Second, they must show that an embodied, generative AI assistant is equal to or surpasses the value of a human instructor in a traditional gross anatomy course. 
    }
    \Description{Virtual reality environment, with two avatars (left: user, right: virtual anatomy assistant) standing in what appears to be a living room around a dissection table with a human cadaver lying on the table. A digital screen is placed behind the table with a learning question.}
    \label{fig:vr-ai-assistant}
\end{figure}

In the modern era, limited cadaver availability~\cite{bodyneedsUGot} as well as reforms in clinical education to espouse student-centered, integrated clinical application curricula~\cite{ghosh2017cadaveric} have resulted in alternative approaches to teaching anatomy beyond the cadaver lab. As a result of this shift, new generations of technology that employ immersive environments and artificial intelligence have been adapted to support medical pedagogy. These new technologies face the same parallel challenges of conveying the data and of making a case for the utility of the method that generated the visual, as did past advances like the X-ray. Chheang et al.'s ~\cite{chheang2023towards} generative AI assistant in a virtual reality anatomy education environment is but one recent example of a new pedagogy of sight (\autoref{fig:vr-ai-assistant}). In this application, the authors need to train and convince a new user base of the validity and utility of virtual reality environments as opposed to traditional presence human cadaver dissection. They must also do the same rhetorical work for the virtual assistant, which is powered by OpenAI's ChatGPT model, to convince users that language models can function as an interactive and adaptive approach to learning anatomy. While specific technologies and data cultures may change, the challenge of building trust in and authority of unfamiliar technology with unknown utility remains.

\section{Reasoning about Disease and Dysfunction}
\label{sec:disease}
In this vignette, we focus on how epistemologies and emerging scientific theories can shape visualizations, in particular those meant not just for contemplation but for pedagogy and practice. The use of anatomical visualization to diagnose and cure the body is inextricably linked with data cultural influences: notions of what is meant by health, what counts as evidence, and how signs and symptoms are structured and collated. Defining disease requires a conception of what a healthy body looks like, which is a normative judgment that has just as much to do with the society in which one is enmeshed as any individual observation.

\begin{figure}
    \centering
    \includegraphics[width=0.8\linewidth]{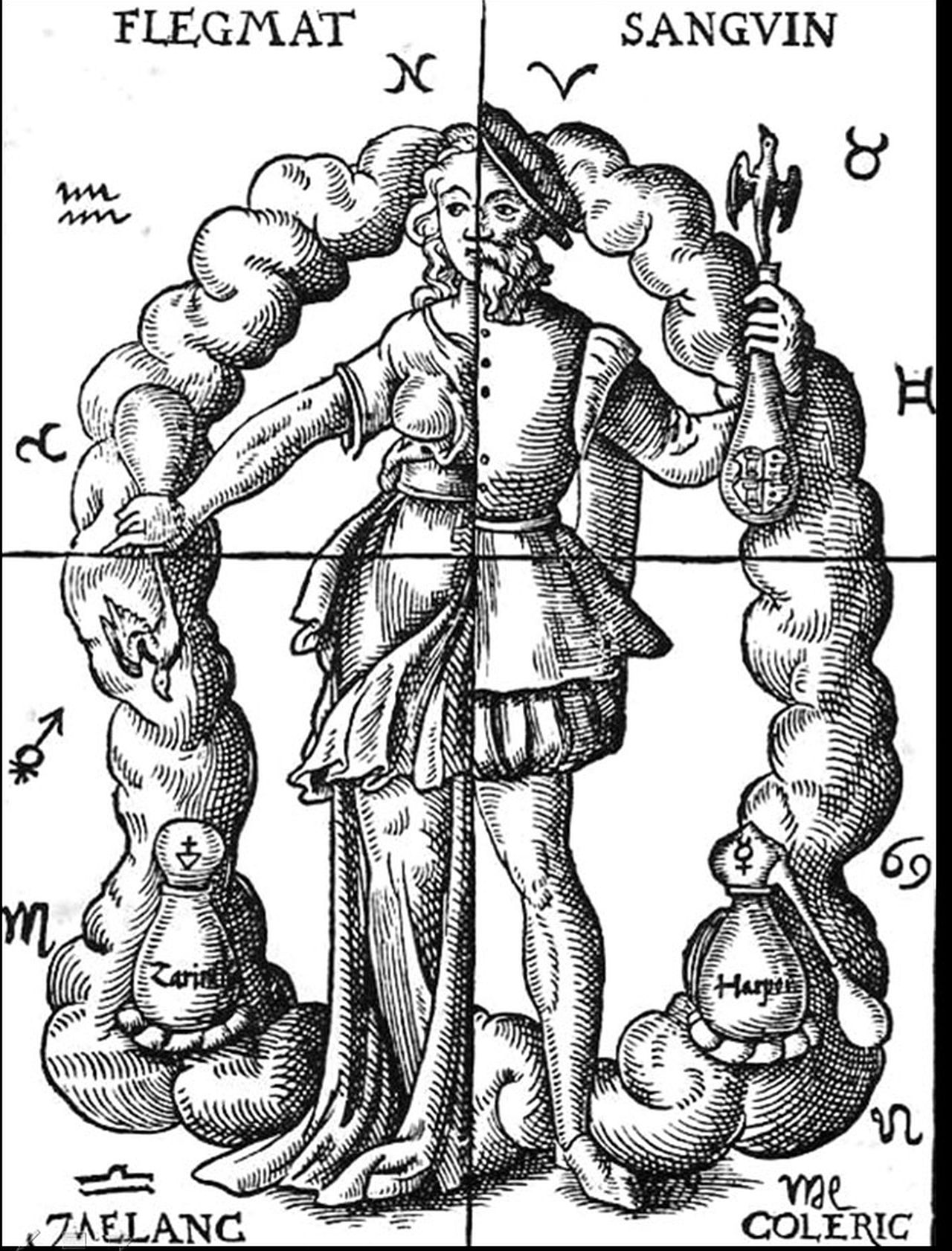}
    \caption{1574 illustration from Leonhart Thurneisser's \worktitle{Quinta Essentia} illustrating the temperaments of a person resulting from excesses of each of the four humours of medieval medicine: Phlegmatic (phlegm), Sanguine (blood), Choleric (yellow bile) and Melancholic (black bile). Surplus of each humour was believed to be a causative agent of disease, but also to have psychological ramifications: in this latter conception, it survived as a classifying tool into 20th century psychology~\protect\cite{eysenck1964principles}, even after the other medical uses had been largely superseded.}
    \Description{Androgynous figure in medieval clothing, divided into four quadrants to represent each of the four humors of medieval medicine.}
    \label{fig:humors}
\end{figure}

Before the acceptance of the germ theory of disease, attribution of various signs and symptoms was driven by other causal theories, which in turn shaped the sort of information that was considered relevant to the physician. For instance, the attribution of diseases to imbalances in the four humours, popularized by early Greek writers like Hippocrates and Galen (but appearing even into the 20th century in the early science of psychology~\cite{eysenck1964principles}), created an entire genre of visualizations that functioned as humoural lookup tables (Figure~\ref{fig:humors}): sorting personality traits or flaws based on their humoural category, and, conversely, going from a detected sign or reported symptom to a humoural cause and then to a suggested remedy: bleeding, purging, or otherwise rebalancing the humours.

\begin{figure}
    \centering
   \begin{subfigure}[b]{0.7\linewidth}
         \centering
         \includegraphics[width=\textwidth]{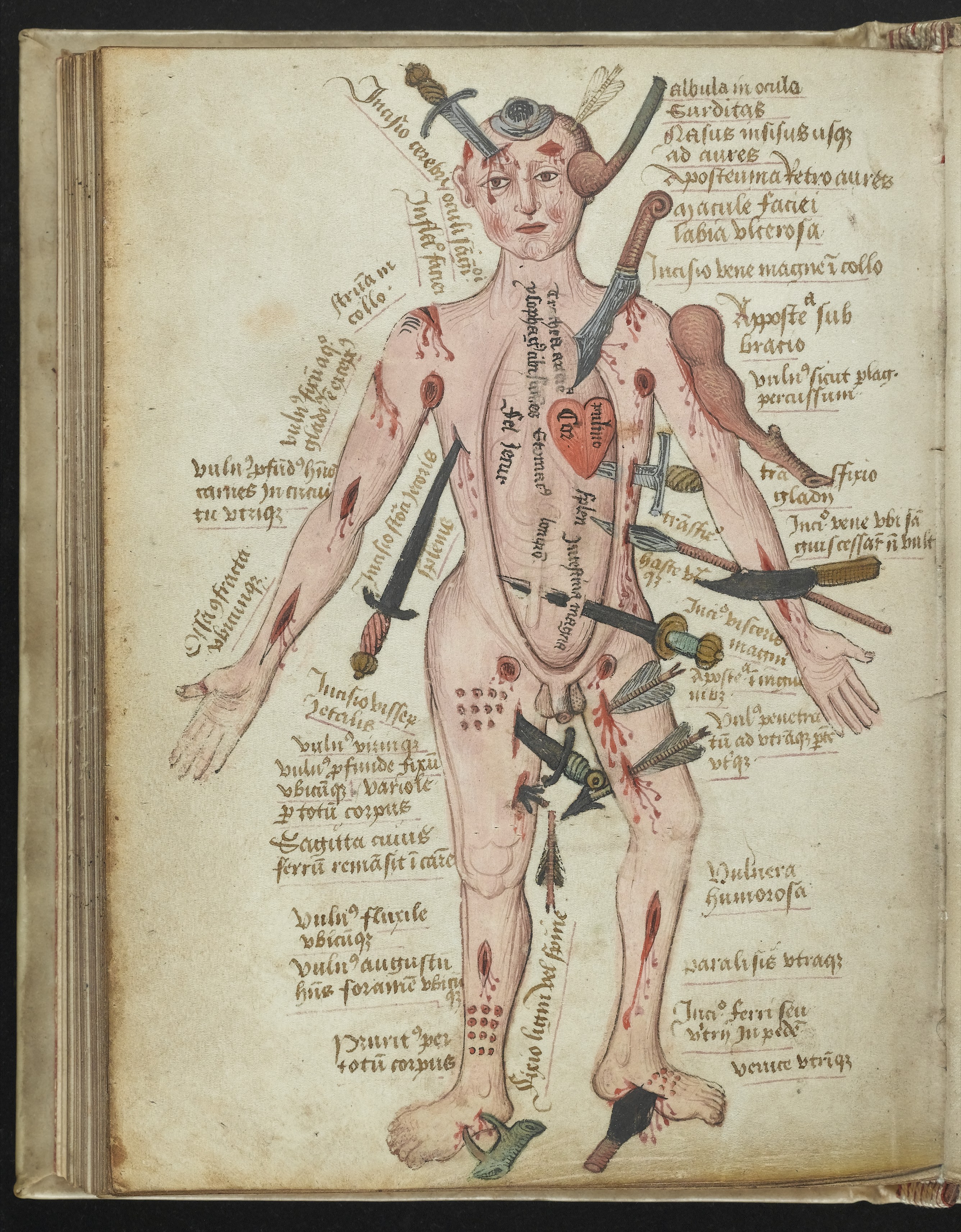}
         \phantomsubcaption
         \Description{Color drawing of a man pierced by swords and other weapons, and bleeding from many simultaneous open wounds. Each wound has a textual label.}
         \label{fig:wound1}
     \end{subfigure}
     \\
     \begin{subfigure}[b]{0.7\linewidth}
         \centering
         \includegraphics[width=\textwidth]{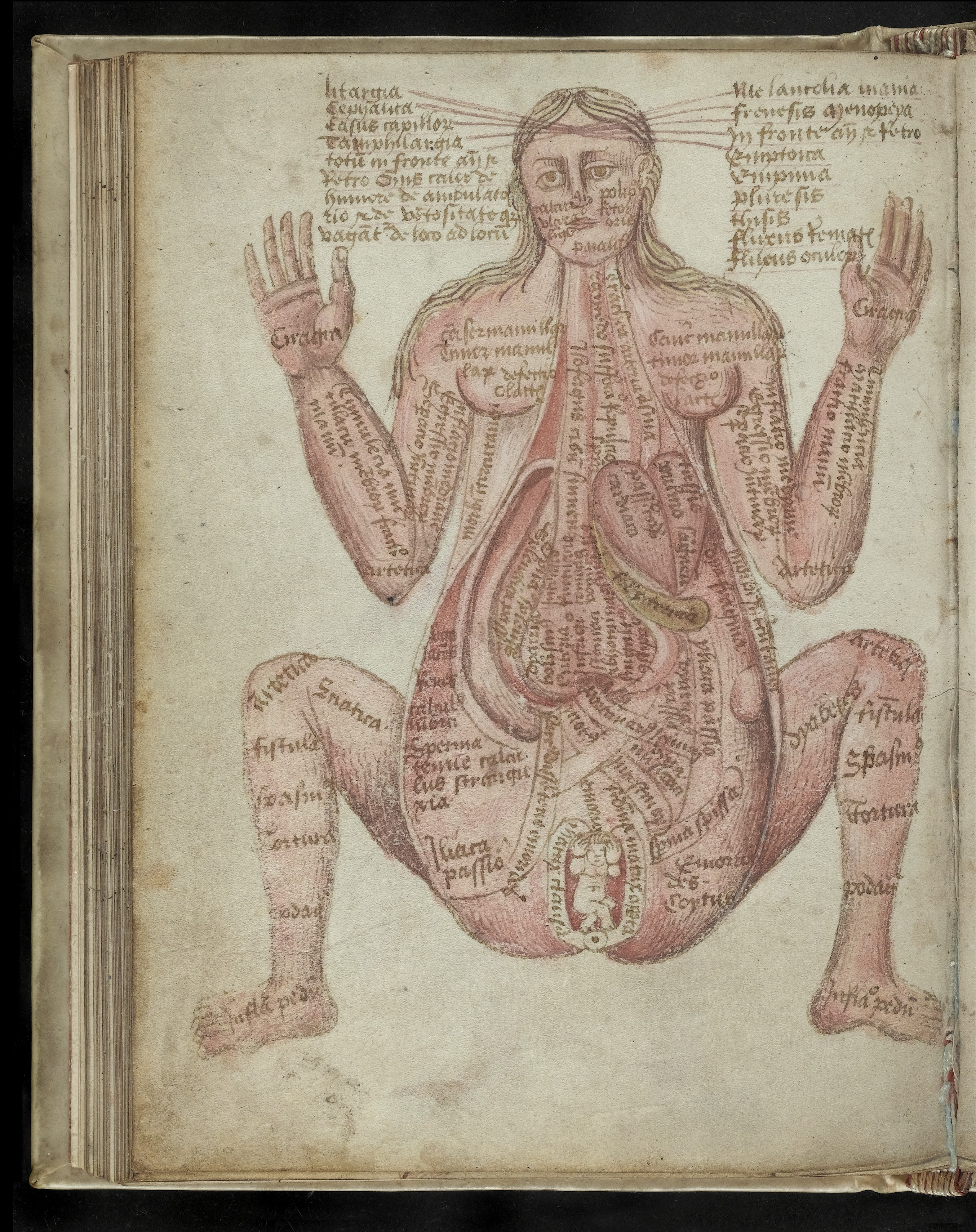}
         \phantomsubcaption
         \Description{Color drawing of a woman covered with writing indicating potential illnesses. Her torso and uterus are expanded to take up the majority of the figure, with labels of illnesses related to childbirth and pregnancy.}
         \label{fig:wound2}
     \end{subfigure}
    \caption{
    A ``wound man'' (top panel) and ``pregnant disease woman'' (bottom panel) from the ca. 15th century English medical manuscript \worktitle{Anatomia}, attributed to ``Pseudo-Galen'' (a complex authorship consisting of both medical texts mis-attributed as Galenic for marketing reasons, as well as texts by other authors that circulated with genuine Galenic texts~\cite{fortuna2020pseudo}). Similar images appeared in many subsequent medical manuscripts, including the late 15th century bundle of medical manuscripts \worktitle{Fasciculus Medicinae} associated with German physician Johannes de Ketham. Note the use of textual annotation; in most wound man drawings, these annotations would be references to the sections of the following manuscript that dealt with the treatment of the specific wound in question, and so the visualizations functioned both as diagrams as well as pictoral indices.
    }
    \label{fig:woundman}
\end{figure}

\begin{figure}
    \centering
    \includegraphics[width=\linewidth]{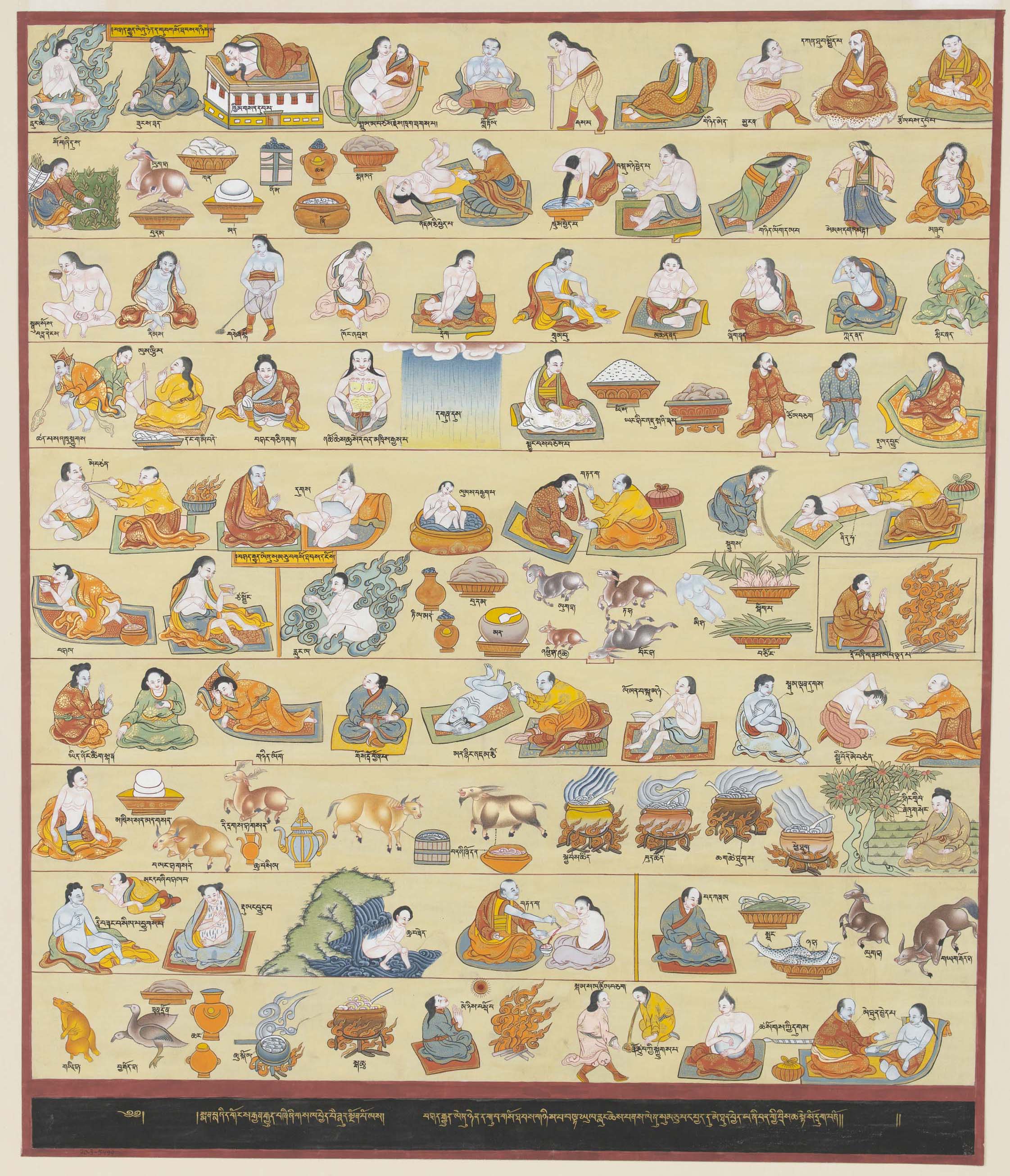}
    \caption{Another \textit{thangka} from the same 17th century \worktitle{Blue Beryl} collection  as \protect\autoref{fig:tibetanatomy} attributed to Desi Sangye Gyatso, here showing a visual encyclopedia of methods of treatment for disease. Note that some figures have blue skin, which can be a visual shortcut for indicating particular imbalances, part of a visual-pedagogical system of ``repetition with a difference'' where distinguishing visual patterns in otherwise similar forms train the viewer to attend to specific pedagogically relevant features~\protect\cite{dachille2012case}.}
    \Description{Rows of illustrated figures juxtaposed with myriad activities, ingredients, or procedures to treat illnesses.}
    \label{fig:tibetandiagnostics}
\end{figure}

The necessity of connecting particular illnesses with (often humoural) causes and treatment led to the creation of a specific offshoot of anatomical visualization: ``human tables''~\cite{hartnell_many_nodate} such as the medieval ``wound man'' and ``disease woman'' where human figures were simultaneously pierced by arrows, stabbed by swords, or otherwise displaying sets of simultaneous diseases, injuries, and disorders (\autoref{fig:woundman}). These arresting figures are annotated with references or other pointers to the relevant parts of the medical manuscripts in which they appeared. In addition to functioning as tables of contents or taking advantage of the ``flexible diagrammatic potential of the human body''~\cite{hartnell2017wording}, these diagrams could also be procedural: for instance, a ``bloodletting man'' associated specific diseases with specific parts of the body to be bled. Other forms of these visual human encyclopedias, such as the aforementioned \textit{thangka} illustrations of medical treatments (\autoref{fig:tibetandiagnostics}), functioned not just as indexes but as aids to memory, using hues and iconic symbols and other forms of ``visual logic''~\cite{dachille2012case} to signal particular illnesses or causes (for instance, blue-colored skin can represent an imbalance in the body's vital energies).

Early cultures of information dissemination made it difficult to precisely integrate chart and text. Medieval wound men and similar diagrams would be generated as separate processes from corresponding medical texts (that would or might circulate independently or be subject to imperfect copying and transcription), producing potential errors. For instance, Hartnell~\cite{hartnell_many_nodate} suggests that a stone-like lesion on the head of a wound man diagram became, through inattention, mistranslation, or error, a non-diagnostic helmet in a later copy of the same diagram. 

These forms of diagrams also had sociopolitical implications. The use of a male figure as the normative or default form of the body relegated women's health and obstetrics to other charts, figures, or manuscripts that rarely circulated in tandem with the manuscripts by (and for) men~\cite{green2000diseases}. The humoural framework underlying the diagrams also had consequences (in later eras, for instance, encouraging ``heroic'' treatments of severe bloodletting and purging~\cite{sullivan1994sanguine}). If the temperament of human beings could be governed by the humours, societies as a whole could also be modeled in similar terms, such as Ottoman schools of economic thought that saw \shortquote{scholars as blood, merchants as yellow and peasants as black bile, and the statesmen or bureaucrats as the phlegm}~\cite{ozveren2015review}. Violent and even genocidal actions have often used the resulting language of contagion or restoration of equilibrium to justify exclusion, expulsion, or destruction of particular subgroups: for instance, victims may be referred to as internal tumors or bacilli to be excised from the political ``body''~\cite{fisk2009words}.

\begin{figure*}
    \centering
    \includegraphics[width=\linewidth]{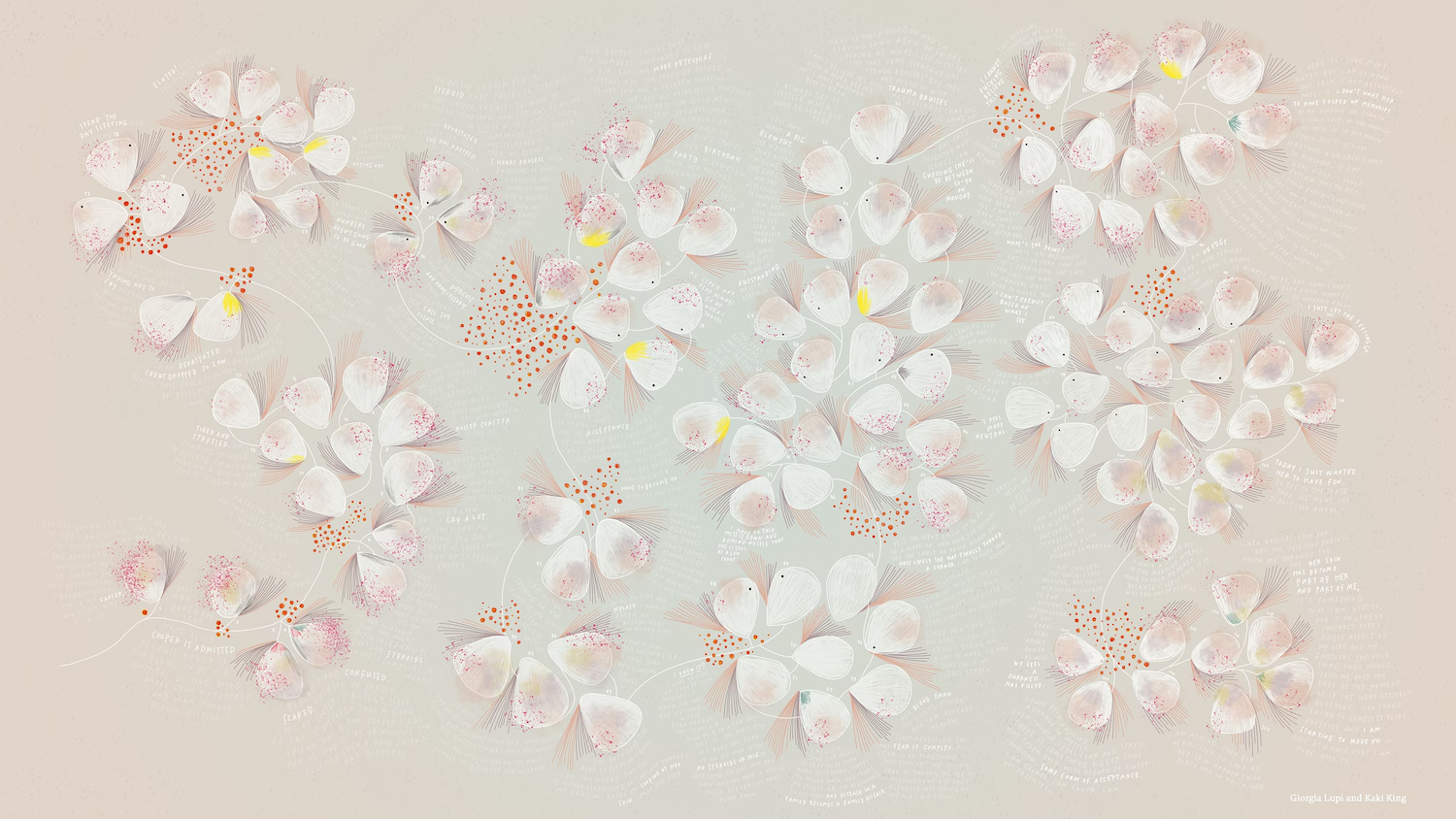}
    \caption{A static overview of Giorgia Lupi and Kaki King's 2018 artwork \worktitle{Bruises: The Data We Don't See}~\protect\cite{lupi2018bruises} showing the course of Kaki King's daughter's illness. Multidimensional glyphs and sounds encode information about symptoms, interspersed with written diary entries and reflections.}
    \Description{Illustration of petal-like glyphs each encoding information about the severity of symptoms, juxtaposed with text excerpts from a symptom diary.}
    \label{fig:bruises}
\end{figure*}

In the modern era, the human-scale visualization of symptoms and signs can take on a different character, as a form of re-individual\-ization or even protest against other forms of data or ways of knowing (as a single diagnosis, as just a medical record, or as just a ``data point''). Slobin~\cite{slobin2014}, when tasked with creating a data visualization showing the variability and severity in Niemann Pick Disease Type C, eventually came to the conclusion that \shortquote{the data visualization is actually people,} and relied on photographs of people with the disease rather than abstract diagrams of multidimensional data. Likewise, King \& Lupi's~\cite{lupi2018bruises} multimedia visualization \textit{Bruises: The Data We Don't See}, documenting the former's daughter's illness, was created partially due to a feeling that \shortquote{clinical records alone hardly capture the impact the illness of a child has on a family} (\autoref{fig:bruises}). The resulting artwork, drawing from photographs and images and medical records, is a multimedia experience that weaves together complex glyphs with animation and music to represent the disease not just as an abstract event but as an emotional experience. 

We contrast the medieval and renaissance use of human-like figures for diagnostics as ways to ``dataify'' unquestioned or untested theories of medical knowledge, with the contemporary use of anthropomorphized or individualized visualizations a way to ``de-dataify'' or otherwise make information that has been abstracted, dispossessed, or otherwise specialized, more human-centered.

\section{The Body as Spectacle}
\label{sec:collaboration}

European Renaissance-era collaborations between anatomists, clinicians, and professional artists blended perspectives and leveraged technologies to produce novel visualizations of the inner human form for a broader audience. Artists trained in drafting, composition, and storytelling produced anatomical visualizations that were more than study aids. These visuals also served as marketing materials showcasing the skills and authority of the scientists and the artists, as commentary on contemporary social and cultural issues, and demonstrated a clear tension between religious and spiritual ideas with the new natural philosophy. Blending art, science, education, and popular culture, increasingly detailed and accurate anatomy of partially-dissected men, women, or even babies, appeared arranged in monstrous, noble, seductive, or comical poses against lush backgrounds of countrysides and architecture~\cite{sappol2006dream}. 

Technological advancements in printing, combined with a world more connected through trade, enabled faster, more facile production and circulation of richly-illustrated anatomical atlases across Europe and beyond to regions still under a dissection moratorium~\cite{akkin2014glimpse,standring2016brief}. Vesalius' \worktitle{De Humani Corporis Fabrica} was in part so successful thanks to its strikingly detailed illustrations, which may have been created by artists from the famous Titian's studio~\cite{kemp1970drawing} (Figure~\ref{fig:musclemen}), and because the publication was so broadly circulated (and liberally plagiarized). 

\begin{figure*}
  \centering
  \begin{subfigure}[b]{0.287\linewidth}
         \centering
         \includegraphics[width=\textwidth]{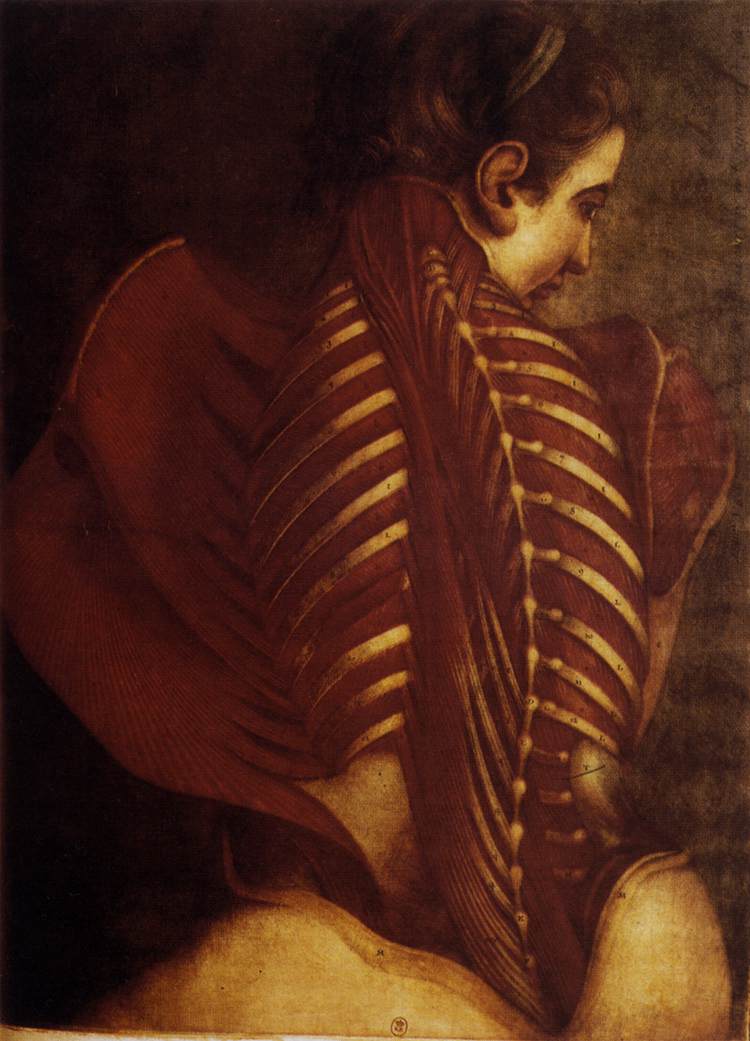}
         \phantomsubcaption
         \Description{Color drawing of a woman. The skin of her back is removed, showing ribs and musculature.}
         \label{fig:gautier}
     \end{subfigure}
  \hfill
 \centering
  \begin{subfigure}[b]{0.6\linewidth}
  \centering
    \includegraphics[width=\linewidth]{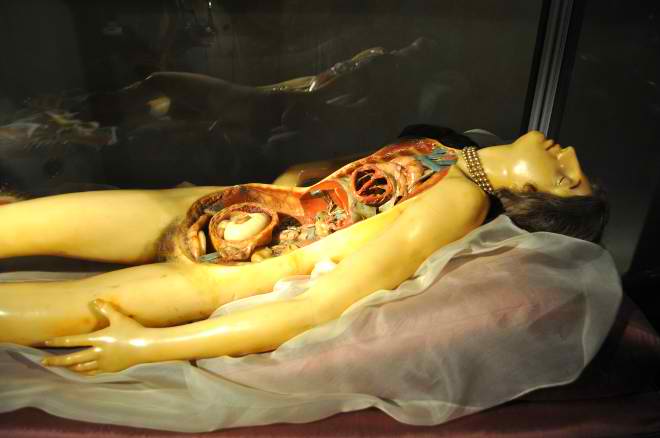}
    \phantomsubcaption
    \Description{Reclining wax figure with her torso cut open, revealing wax internal organs.}
    \label{fig:medicivenus}  
  \end{subfigure}
  \caption{
  \textit{Left panel:} Gautier-d'Agoty's image of a woman, informally called the \worktitle{Flayed Angel} from his 1746 \textit{Myologie Complete} is one of first examples of color in medical printing, showing an elegant but anatomized woman~\cite{sappol2006dream}. 
  \textit{Right panel:} \worktitle{Venerina} or \worktitle{Little Venus} anatomical wax model by Clemente Susini, currently on display at the Museo di Palazzo Poggi in Italy, is one of many life-size dissectible wax models created in the 1780s for both study and public engagement.  
  }
  \label{fig:venuses}
\end{figure*}

Visual representations of the body extended beyond two dimensions and adapted to new needs and technologies. Dissectable wax models like the 18th-century \worktitle{Anatomical Venus} were created to nullify the need for human dissection with a beautiful, doll-like, anatomically accurate instructional model (Figure~\ref{fig:venuses}, right panel). Popular for teaching and for public outreach, such models produced in this period made, to the modern eye, a bizarre balance between art and science, entertainment and education, exploitation and reverence~\cite{ebenstein2016anatomical}, that are echoed in modern plastinated human body exhibits like \worktitle{Body Worlds}~\cite{bodyworlds}, with similarly contentious results. Per Fontana, the first director of the wax model workshop known as La Specola: 
\begin{quote}
    \textit{If we succeeded in reproducing in wax all the marvels of our animal machine, we would no longer need to conduct dissections, and students, physicians, surgeons, and artists would be able to find their desired models in a permanent, odour-free, and incorruptible state. }
\end{quote}

The use of the female form, inspired by the goddess Venus, shows an effort to nudge anatomy from a study of the macabre to that of beauty to appeal to a broader audience, similar to earlier color mezzotint paintings like d'Agoty's \worktitle{Flayed Angel} (Figure ~\ref{fig:venuses}, left panel). The use of wax for modeling gives a life-like sense to the model, who is posed as if resting, nude apart from a pearl necklace. The model can be dissected down seven layers, with the final layer revealing a small fetus in her uterus~\cite{ebenstein2016anatomical}. Anatomical depictions of women often showed them as pregnant (Figures~\ref{fig:woundman} and \ref{fig:mansur}, bottom panel)---a reflection of the role that women, or rather their bodies, served in society, and of the focus in medicine for women on the (often hazardous) processes of pregnancy and birth. In a period where nudity in Europe was controlled, models like the \worktitle{Venus} traveled widely in fairs and museum exhibits and legitimized public engagement with a naked female body, leveraging the lurid or the shocking rather than the purely clinical. Per Sappol, a medical historian:
\begin{quote}
    \textit{Public discussion of sexual desire, ostensibly part of the medical technology of self-regulation, provided an opening through which the pleasure principle could be smuggled in, a fact that popular anatomists were well aware of, and manipulated to their benefit}~\cite{sappol2004morbid}.
\end{quote}

\begin{figure}
 \centering
\includegraphics[width=0.9\linewidth]{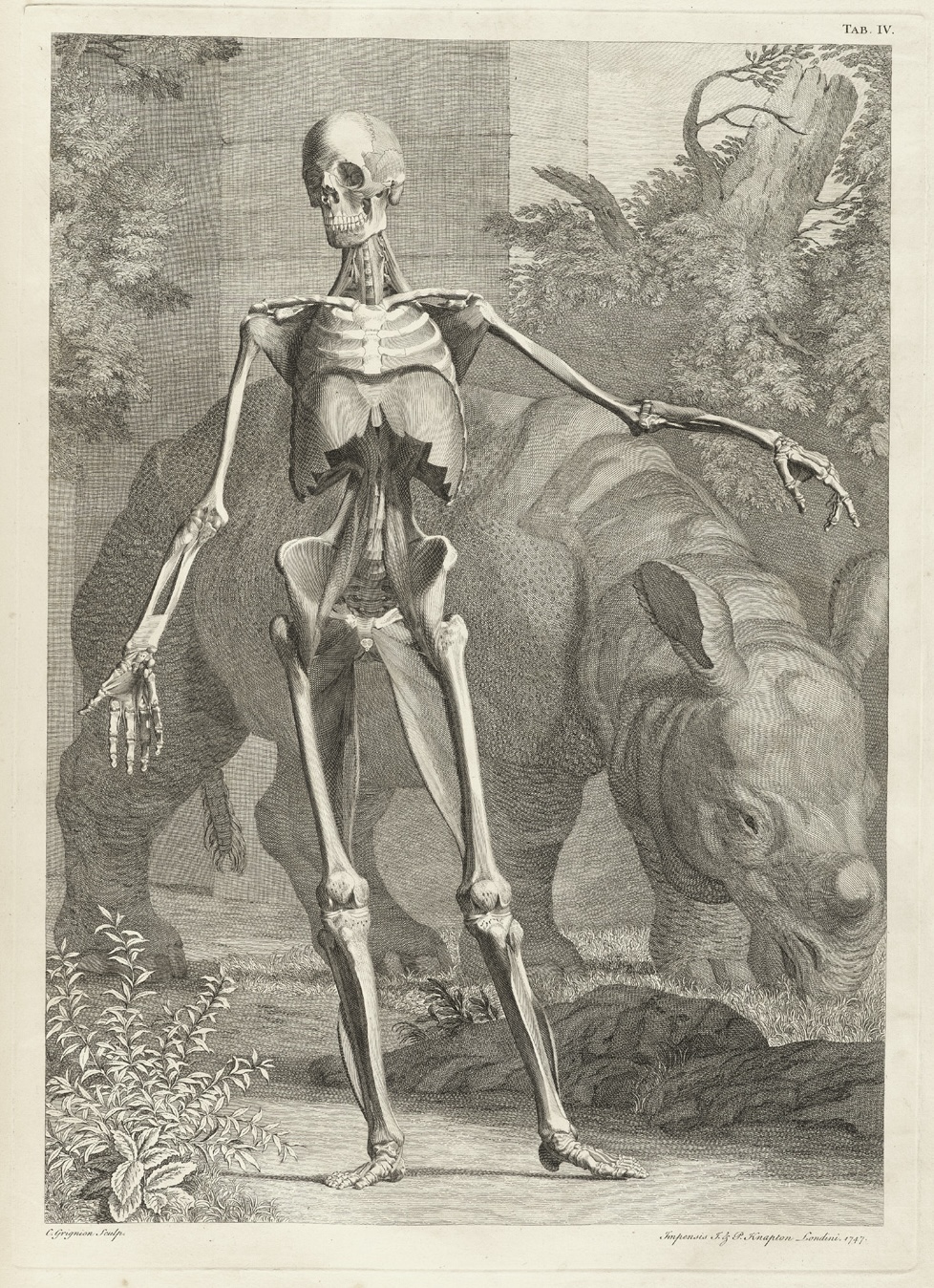}
  \caption{Anatomist and artist teams of 1700s Europe showcase mezzotint printing technologies resulting in visualizations that blur line between art and science, education and social commentary, incorporating popular culture as well as anatomy. This anatomical visualization from Albinus and Wandelaar's 1747 \textit{Tabulae sceleti et musculorum corporis humani} (``of the male sex, of a middle stature and very well proportioned; of the most perfect kind, without any blemish or deformity''~\protect\cite{daston1992image}) includes the famous (at the time) Clara the rhinoceros.
  }
  \Description{Drawing of a skeleton posing contrapposto in front of a rhinoceros.}
  \label{fig:albinus}
\end{figure}

As printing techniques continued to improve in quality and detail, artists continued pushing the bounds of possibility in collaboration with anatomists to connect with a broadening public audience that often came at the expense of clear communication of anatomical information, at times incorporating popular culture references, such as the 18th century anatomical illustration with Clara the rhinoceros in the background (Figure ~\ref{fig:albinus}), who was part of a traveling zoo in Europe around at the time~\cite{ruiz2016tedmed, sappol2006dream}. Per the anatomist Albinus, \shortquote{We thought the rarity of the beast would render these figures of it more agreeable than any other ornament}~\cite{daston1992image}.


\begin{figure*}
   \centering
   \begin{subfigure}[b]{0.32\linewidth}
         \centering
         \includegraphics[width=\linewidth]{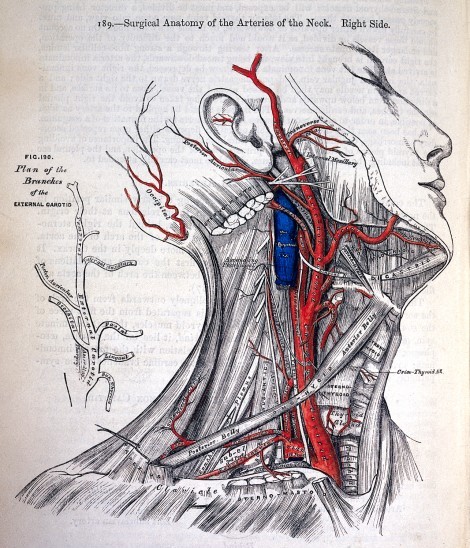}
         \phantomsubcaption
         \Description{Illustration of the neck with most structures in black and white, but with the arteries in strong red, with text labels running their length.}
         \label{fig:gray}
     \end{subfigure}
     \hfill
     \begin{subfigure}[b]{0.53\linewidth}
         \centering
         \includegraphics[width=\linewidth]{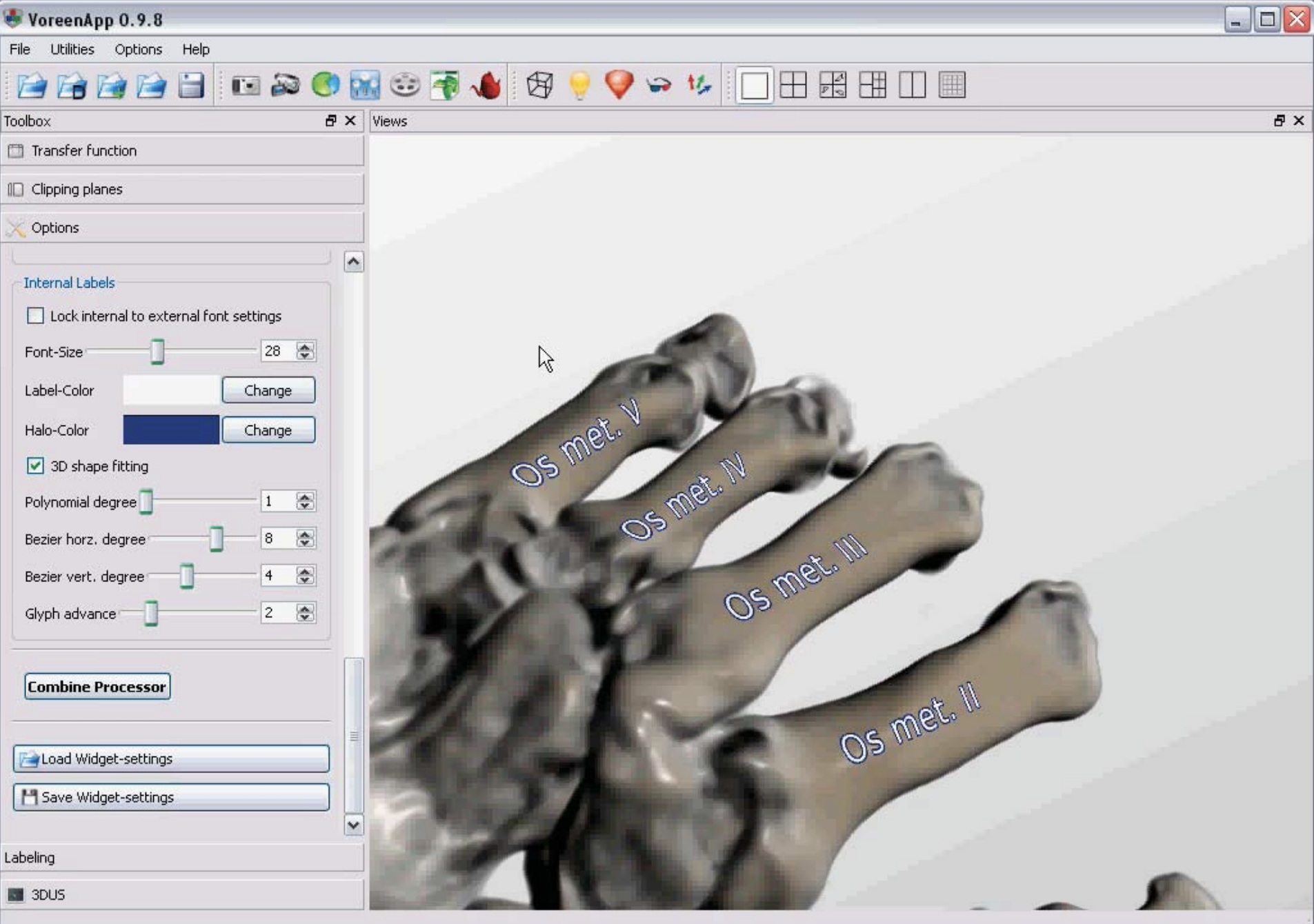}
         \phantomsubcaption
         \Description{3D colored render of the hand, with major bones labeled in situ.}
         \label{fig:ropinski}
     \end{subfigure}
    \caption{\textit{Left panel:} Surgical anatomy of the arteries of the head and neck, \worktitle{Gray's Anatomy} (1858), illustrated by Henry V. Carter. Note selective coloring of relevant structures, and direct annotation of structural anatomy. 
    \textit{Right panel:} Shape and 3D surface fitting of anatomical annotations onto an anatomical surface model derived from a segmented medical dataset~\cite{RopinskiPRH07}.}
    \label{fig:grayropinski}
\end{figure*}

Increased circulation and demand with growing numbers of medical schools offering formal training in medicine created sufficient market pressures in the 19th and 20th centuries for a new profession of artists trained both in the arts and medicine to create illustrations that leveraged storytelling strategies for purely didactic, rather than social commentary, purposes~\cite{sappol2006dream}. 
These medical illustrators further leveraged the possibilities of color printing to selectively color structures of interest for increased visual salience, and began to integrate anatomical structure labels directly into the visualizations as in the iconic \worktitle{Gray's Anatomy} (Figure~\ref{fig:grayropinski}, left panel) rather than using separate legends (as in Figure~\ref{fig:musclemen}) that required the eye to travel back and forth between pages. These storytelling approaches are the bedrock of medical illustration and visualization today, with contemporary visualization research exploring methods to stylistically automate these techniques (Figure~\ref{fig:grayropinski}, right panel)~\cite{cipriano2008text,RopinskiPRH07}.

Modern medical visualization now features collaborations with not only anatomists, clinicians, and artists, but also computer scientists. 
These contemporary approaches often derive from imaging data rather than direct observation, introducing the possibility of illustrative-style visualizations customizable to individual patient anatomy~\cite{preim2013visual}, rather than a generalized or idealized anatomical perspective necessitated by traditional medical illustration~\cite{garrison2021bioviscom}. Additionally, in echoes of the artistic anatomical heyday of the European Renaissance and Enlightenment, contemporary artists are exploring visual representations of the inner body to support personal, social, and political rhetorical agendas in new media: through street art, body paint, and comics~\cite{ruiz2016tedmed}.

\section{Representation of the Marginalized}
\label{sec:ethics}

Throughout its history, medical visualization has overwhelmingly relied on the bodies of the marginalized of society, producing a disconnect between the human source and the resulting visualizations that are most often heroic, white, and male, even into today. Convicted criminals were, for centuries, the only legal possibility for anatomists and clinicians to obtain bodies for dissection~\cite{zimmer2005soul}, with supply pressures creating a black market for fresh corpses through ``Resurrection Men'' who trawled graves of the poor (and in the United States often Black) deceased whose families lacked the money and power to prevent these activities well into the nineteenth or even twentieth centuries~\cite{davidson2007resurrection,halperin2007poor,hutton2015study}. The dissection of criminals and other marginalized groups unavoidably implies a certain moralistic code and reinforces societal structures of power, denying these bodies the sanctity normally bestowed on the human body while using them as a means to further scientific discovery. Threads between criminality and the acceptability of human dissection persist in contemporary society. A recent survey spanning 68 countries using cadavers in undergraduate medical curricula found that only 32\% of sourced bodies are explicitly donated---the remaining are ``unclaimed'', and one country notes that \shortquote{the bodies of executed prisoners are given to anatomy departments}~\cite{habicht2018bodies}. The U.S. National Library of Medicine's \worktitle{Visible Human Project}~\cite{ackerman2000imaging}, released in 1994, is a publicly-available, high-resolution dataset of the complete human body used in thousands of research, artistic, educational, and clinical applications. One of the cadavers in the \worktitle{Project} is a convicted murderer, executed for his crimes and who donated his body to science. Similarly, the aforementioned \worktitle{Body Worlds} exhibit~\cite{bodyworlds}, which uses plastinated cadavers, has been a subject of controversy over the source of its ``exhibits'' (including claims that some of the bodies were from non-consenting prisoners or mental patients) and whether the artistic or pedagogical benefits of the exhibit overcomes the ethical and moral issues associated with the public display of private bodies~\cite{burns2007gunther}.

As a particularly troubling modern example, the \worktitle{Pernkopf Atlas} (1943) is recognized for its beautiful, lifelike rendering of human tissue, but the ethics surrounding the \worktitle{Atlas} are suspect: based in Vienna during the rise of Fascism and World War II, Pernkopf and his team of illustrators were members of the Nazi Party. Circumstantial, although not definitive, evidence suggests that the bodies dissected and illustrated for the \worktitle{Atlas} were Holocaust victims~\cite{atlas2001ethics}. Debates have raged in the medical and medical-adjacent communities on the ethics of using, let alone discussing, the \worktitle{Atlas} in light of this information~\cite{riggs1998should}. For the anatomists conducting these early-day dissections or the modern-day student reading the \worktitle{Pernkopf Atlas}, accessing the \worktitle{Visible Human}, or visiting \worktitle{Body Worlds} to study anatomy: did/do they reflect on the life and actions of the once-living body in front of them, or do they see only viscera? 

Issues of consent and representation in anatomical visualizations extend to the living. Prior to the invention and broader adoption of imaging methods, clinicians' sole possibility to examine the appearance and functionality anatomical structures of \textit{live} human subjects was, essentially, through patients wounded in battle or who suffered chance accidents. The ethics surrounding clinical relationships with such patients at this time was questionable at best, based as these relationships were on \shortquote{models of employment, servitude, and labor}~\cite{green2010working}. Alexis St. Martin in 1822 received a gunshot wound that, in a strange twist of the healing process, left an open window to his stomach, what is medically termed a \textit{gastric fistula}. St. Martin's doctor, William Beaumont, seized upon this opportunity for discovery: he contracted the illiterate and now-destitute St. Martin as his servant, and for the next several years conducted a series of experiments to elucidate the properties of the stomach and digestion, taking copious notes and illustrations of the results with a degree of detachment that the reader nearly forgets that this was a live human subject~\cite{beaumont1838experiments}.

As discussed earlier with the \worktitle{Anatomical Venus} and the \worktitle{Flayed Angel} (\autoref{fig:venuses}), there are ethical issues around the objectification of the body (especially the female body), and the extent to which prurient or voyeuristic interests clash with the clinical or pedagogical. In 1971, the \textit{Anatomical Basis of Medical Practice}, targeted to first-year medical students, went to press with suggestive commentary and photographs of women in \worktitle{Playboy}- or ``pin-up''-style poses to motivate a (presumably) lonely male student to learn surface anatomy. Its publication caused a huge and immediate backlash concerning the perceived objectification and denigration of women and was quickly pulled, but not before becoming an icon of the women's movement and being used in debates on how to define pornography~\cite{halperin2009pornographic}, echoing contemporary disquiet with the use of the \worktitle{Lena} image from \worktitle{Playboy} in contemporary computer graphics~\cite{munson1996note}. The selective use of attractive and suggestive women introduces another aspect of contemporary representation---an overwhelming majority of anatomical visualizations depict ultra-fit, white, male bodies that are not reflective of the diversity of shape, color, and gender of society. The result is that clinicians are not well-prepared to deal with diagnoses of, for example, skin conditions on non-white skin, and that BIPOC and non-binary patients cannot see themselves in the canon of Western medicine. Although there are growing efforts to motivate change~\cite{amidiversityfellow}, the road is still long to equal representation~\cite{viralbipocanatomy}.  

These examples surface difficult bioethical questions extending far beyond the medical domain on the use of scientific data colored by their associations. They suggest that collection of data about the body is not neutral, but is influenced by biopolitical concerns (or even ``necropolitical''~\cite{mbembe2003necropolitics} concerns---the politics over how people are allowed to live or die). Can, or should, data be assessed on their own merits, excluding the biopolitical contexts in which they were acquired, or is context inextricably linked to the ultimate---in this case visual---product? By engaging with visualizations built from data with bioethically troubling origins, are we showing tacit support for racism, sexism, Nazism, and so forth, or can we see this as an opportunity to honor the dead with reflective, considered engagement with these works? We argue that context is impossible to excise from data and that, while these discussions are thorny, they are imperative to hold. Avoidance of these bioethical debates is the greatest disrespect we could show the living or the dead. 

\section{Discussion}

The history of medical visualization is a long and fragmented story, capturing only snapshots and partial images while the majority are lost to time. These limited views, however, can provide a window into the data cultures of the period in which a medical visualization was produced, lending insight to the historical, technological, social, and political contexts that are inextricably tied to the resulting visualization and its mark on society. In this paper, we focused on these paradigm-shifting issues that produced the conditions to enable so-called groundbreaking medical visualizations such as the Vesalian muscle-men or \worktitle{Gray's Anatomy}. Turning now from our review of the past and present to look toward the future, we discuss select lessons learned from history and reflect on these in current and future medical visualization practice. 

We first observe how visual forms and genres, rather than appearing suddenly and uniquely, instead periodically recur in new guises or for different aims, changed or recapitulated by new cultural or epistemic contexts. As Marx~\cite{marx1926eighteenth} famously states, people: 
 \begin{quote}
    \textit{...make their own history, but they do not make it as they please; they do not make it under self-selected circumstances, but under circumstances existing already, given and transmitted from the past. The tradition of all dead generations weighs like a nightmare on the brains of the living.}
 \end{quote}
The graph-like iconographic ``subway map'' forms of anatomical drawings, made of necessity in the face of taboos on dissection, appear again in the guise of modern simplified and abstract medical illustrations meant for pedagogy and dissemination: the children's game of \worktitle{Operation} (Figure \ref{fig:modernschema}, right panel) and the children's character Slim Goodbody are not entirely dissimilar from these early diagrams, and for not entirely dissimilar reasons! The idealized forms of the body in later anatomical drawings, meant to provoke interest and elide the often messy nature of anatomy, recur again in the use of pin-ups to illustrate surface anatomy. The close integration of text and image occurs in the works of Man\d{s}\={u}r ibn Ily\={a}s, Vesalius, Gray, and in modern 3D renders. What changes, then, are the material and productive forces underpinning these visual forms, the aforementioned \textit{data cultures} that shape what data is counted and what data counts. Researchers and practitioners of visualizations must be mindful of these data cultures, or face potential obscurity or irrelevance. This mindfulness extends also to the ethical and political ramifications of what data we collect and how we visualize it. Questions of the ethics of dissection and how our bodies are used after our death did not end in Alexandria, or with the hangings of graverobbers like Burke and Hare, but continue in exhibits like the \worktitle{Visible Human Project} or \worktitle{Body Worlds}, or with the use of Henrietta Lacks' cell lines in cancer research. The politics of who is counted (and who does the counting) is not limited to the dissection table, but is core to projects like \worktitle{Data Feminism}~\cite{d2020data}.

Connected to our observations of recurring visual genres is our observation of dialectical processes of ``datafication'' and ``de-datafication'' that seem to ebb and flow across cultural contexts. Taboos against dissection, and the perceived sanctity or inviolability of the human body, for a time kept anatomical information limited. Not just technological but \textit{cultural} and \textit{rhetorical} processes changed this status quo and made the body licit for analysis. This process involves a necessary stage of teaching people how to interpret these new sources of data, ``selling'' the value or utility of these new source of data, first to experts or authorities but then to mass audiences. When these data are commonplace, there are then reactions to once again re-sanctify or otherwise re-personalize them. The emerging genres of anthropographics~\cite{morais2020showing} are modern examples of reactions in this space, as are works like the \worktitle{New York Times}'~\cite{nyt} \worktitle{How America Lost One Million People} that switch between ``close'' (personal stories and anecdotes) and ``distant'' (histograms, thematic maps, and other statistical graphics) data presentations in an attempt to both convey at scale but also re-personalize phenomena 
that had become too abstract or data-like.

Our last class of observations concerns the primacy of the \textit{personal} and the \textit{political} in medical visualization. Haraway introduced the concept of \textit{situated knowledge}~\cite{haraway1988situated} to argue that knowledge production is partial and perspectival, taking place under specific historical, political, and situational circumstances. Contemporary visualization and HCI research recognizes the importance of situated knowledge through practices such as design study methodology~\cite{meyer2019criteria} and co-creation workshops. In the context of medical visualization, history correspondingly shows that there are no one-size-fits-all objective design solutions when mapping the structures and diseases of the body. Just as Renaissance-era visualizations of anatomy blended theology with direct observation in a reflection of society at that time, so should current and future designers consider today's climate of science skepticism and fake news. Pedagogies of sight developed by the likes of Vesalius, Hooke, and Ram\'{o}n y Cajal in centuries past were products of their particular environments, and these pedagogies must be continually revisited and revised to instruct and convince new generations. 

Personalized, precision medicine is the byline of today's diagnostic medical practice, calling for tailored approaches and visualizations specific to the individual. For instance, our current era of digitalization and datafication combined with the COVID-19 pandemic has brought epidemiological information (and the resulting visualizations and ``counter-visualizations''~\cite{lee2021viral}) in unprecedented detail to anyone who may be interested. However, to what extent is this detail necessary? More broadly speaking, what kind of information does a person truly need to access, and can we share this information in a way that it resonates, is ethical, and is accessible? Recognition of knowledge as situated, and the replacement of \textit{given} ``data'' with \textit{taken} ``capta''~\cite{drucker2011humanities} may help to guard against biases leading to negative health outcomes on the part of the patient and physician. Consider the treatment and management of pain: numerous clinical studies have documented differences in pain perception along ethnic and gender lines~\cite{campbell2012ethnic}. Explanations for these differences are contentious, complex, and multifaceted but have revealed shocking implicit biases in the healthcare system that often result in physicians under-assessing and consequently under-treating pain in women and ethnic minority groups. This belies claims that differences in pain experiences are due to biological differences alone~\cite{hoffman2016racial,bostick2018pain}. Designing visualizations or visual analytics applications intended to support clinical decision-making and appropriate communication requires situated knowledge to address and (attempt to) course-correct when a data culture politicizes or depersonalizes the lived experiences of individuals.





\section{A Call to Action}
Our observations from medical visualization lead us to suggest reforms in how we ought to operate as designers, researchers, or circulators of visualizations more broadly. 

The first is to \textbf{remember our history}, rather than to discard historical visualizations as relics of forgotten or misguided eras. Designers of these visualizations faced many problems that remain today: dealing with uncertainty and the unknown, conveying complex information to mass publics, or convincing audiences of the utility or accuracy of new or unfamiliar sources of information. That the designers of these illustrations and visualizations came from temporally, geographically, or (even) intellectually remote places should not disqualify them as sources for inspiration or reference.

Secondly, we suggest the need for a more prominent \textbf{close analysis of visualizations as rhetorical objects}. Much as literary scholars can use close reading to examine the use of language and meter in poems, we believe that a similar \shortquote{close reading for visualization}~\cite{janicke2015close} could provide richer information than just a study of usability or assessment of visual clarity. We invite the reader to take time to contemplate our examples with an eye towards exploring how the visual design creates meaning, and how the final form of the graphic is or is not contingent on a particular data culture. Quantitative evaluations of many of these visualizations would, in many cases, miss the point: how effectively one could extract information about where one ought to bleed a patient to balance their humours may utterly fail to capture the utility (or non-utility) of a diagram. Assessments of the Tufte-style data ink implications of including, say, an extraneous famous rhinoceros in an anatomical drawing may fail to recognize how such decisions were efforts to popularize or regularize information, rather than simply communicate an anatomical fact.

We also call on the field to \textbf{consider visualizations in context}, rather than to view them as stand-alone artifacts. The visualizations we present all have the same object of inquiry: the human body. Yet they are by no means interchangeable. Many are even largely illegible without knowledge of the underlying theories (or theologies) of bodies and health that under-girded their creation. The visualizations and illustrations that we produce today are no different, and it is likely that a student of visualization in the future might view our own efforts with the same sort of amused disbelief with which we might contemplate ancient illustrations built on discredited medical theories like humourism. Visualizations are highly \textit{contingent} artifacts: they depend on a host of factors that are exterior to the page or screen in which they are viewed. Designing, assessing, or comparing visualizations must take these contingencies into account. 

The full context of a visualization includes a \textbf{consideration of the ethical and moral components of data collection and visualization}: inequalities of gender, race, class, and ability are an inescapable part of how bodies become data, and how those data are visualized. Issues of consent, representation, and equity are perhaps most graphically evident when a literal human is (or is not) depicted. However, these issues occur in other datasets as well, and our responsibilities and duties to other people apply to every visualization we create~\cite{correll2019ethical}.

\section{Conclusion}
Changing sociocultural perspectives, epistemologies, and technologies have motivated new ways of seeing and visually representing the human body. What is surfaced in illustrations and visualizations is a product of what we are trained to see, and what is valued in our practices. Philosophies of the body vary across cultures, eras, and individuals, and influence the resulting anatomical diagrams. 

Today, pregnancies are shared and celebrated by sending pictures of sonograms to group chats. Doctors will point to X-rays or biopsy results when communicating with their patients. Advertisements for diets, medication, or supplements will show rendered images of muscles and tissue, arteries and lungs. This omnipresence of anatomical visualizations is part of the superstructure of a data culture that has very specific epistemic conceptions of information and objectivity, that builds off a centuries-long empirical and technological project towards systematizing the structures of the body, and an equally long project of using persuasion and communication to convince audiences of the utility of all of this resulting data.


Much as one may like to believe that history traces a straight line of continuous novelty and innovation, our investigation into anatomical visualization found echoes of visual forms and genres across space and time, repurposed to serve new agendas. Visualizations of anatomy, imaging, or disease data reflect their contemporary sociopolitical milieux, and are only the latest in a long historical lineage of attempts at visual sense-making. These currents are unstable: the cultures of data, production, and circulation were different in Galen's time, just as they were for Gray, and will be different for the next generation of visualization designers. Understanding a visualization, let alone authoring a successful one, requires an awareness of these sociopolitical currents.

\begin{acks}
We would like to thank Miriah Meyer, David Grellscheid, and Laura Felley for their comments on drafts of this work. 
This work is supported in part by the University of Bergen and the Trond Mohn Foundation in Bergen (\#813558, Visualizing Data Science for Large Scale Hypothesis Management in Imaging Biomarker Discovery (VIDI) Project). 
\end{acks}

\bibliographystyle{ACM-Reference-Format}
\bibliography{sources}


\begin{thebibliography}{104}


\ifx \showCODEN    \undefined \def \showCODEN     #1{\unskip}     \fi
\ifx \showDOI      \undefined \def \showDOI       #1{#1}\fi
\ifx \showISBNx    \undefined \def \showISBNx     #1{\unskip}     \fi
\ifx \showISBNxiii \undefined \def \showISBNxiii  #1{\unskip}     \fi
\ifx \showISSN     \undefined \def \showISSN      #1{\unskip}     \fi
\ifx \showLCCN     \undefined \def \showLCCN      #1{\unskip}     \fi
\ifx \shownote     \undefined \def \shownote      #1{#1}          \fi
\ifx \showarticletitle \undefined \def \showarticletitle #1{#1}   \fi
\ifx \showURL      \undefined \def \showURL       {\relax}        \fi
\providecommand\bibfield[2]{#2}
\providecommand\bibinfo[2]{#2}
\providecommand\natexlab[1]{#1}
\providecommand\showeprint[2][]{arXiv:#2}

\bibitem[Ackerman and Banvard(2000)]%
        {ackerman2000imaging}
\bibfield{author}{\bibinfo{person}{MJ Ackerman} {and} \bibinfo{person}{RA
  Banvard}.} \bibinfo{year}{2000}\natexlab{}.
\newblock \showarticletitle{Imaging outcomes from the National Library of
  Medicine's Visible Human Project}.
\newblock \bibinfo{journal}{\emph{Computerized medical imaging and graphics:
  the official journal of the Computerized Medical Imaging Society}}
  \bibinfo{volume}{24}, \bibinfo{number}{3} (\bibinfo{year}{2000}),
  \bibinfo{pages}{125--126}.
\newblock


\bibitem[Akkin and Dinc(2014)]%
        {akkin2014glimpse}
\bibfield{author}{\bibinfo{person}{Salih~Murat Akkin} {and}
  \bibinfo{person}{Gulten Dinc}.} \bibinfo{year}{2014}\natexlab{}.
\newblock \showarticletitle{A glimpse into the process of gaining permission
  for the educational dissection of human cadavers in the Ottoman Empire}.
\newblock \bibinfo{journal}{\emph{Clinical Anatomy}} \bibinfo{volume}{27},
  \bibinfo{number}{7} (\bibinfo{year}{2014}), \bibinfo{pages}{964--971}.
\newblock
\urldef\tempurl%
\url{https://doi.org/10.1002/ca.22421}
\showDOI{\tempurl}


\bibitem[Aristotle(2004)]%
        {aristotle}
\bibfield{author}{\bibinfo{person}{Aristotle}.}
  \bibinfo{year}{2004}\natexlab{}.
\newblock \bibinfo{booktitle}{\emph{Rhetoric}}.
\newblock \bibinfo{publisher}{Dover}.
\newblock


\bibitem[Atlas(2001)]%
        {atlas2001ethics}
\bibfield{author}{\bibinfo{person}{Michel~C Atlas}.}
  \bibinfo{year}{2001}\natexlab{}.
\newblock \showarticletitle{Ethics and access to teaching materials in the
  medical library: the case of the Pernkopf atlas}.
\newblock \bibinfo{journal}{\emph{Bulletin of the Medical Library Association}}
  \bibinfo{volume}{89}, \bibinfo{number}{1} (\bibinfo{year}{2001}),
  \bibinfo{pages}{51}.
\newblock


\bibitem[Beaumont(1838)]%
        {beaumont1838experiments}
\bibfield{author}{\bibinfo{person}{William Beaumont}.}
  \bibinfo{year}{1838}\natexlab{}.
\newblock \bibinfo{booktitle}{\emph{Experiments and Observations on the Gastric
  Juice, and the Physiology of Digestion}}.
\newblock \bibinfo{publisher}{MacLachland \& Stewart}.
\newblock


\bibitem[Benjamin(1999)]%
        {benjamin1999arcades}
\bibfield{author}{\bibinfo{person}{Walter Benjamin}.}
  \bibinfo{year}{1999}\natexlab{}.
\newblock \bibinfo{booktitle}{\emph{The Arcades Project}}.
\newblock \bibinfo{publisher}{Harvard University Press}.
\newblock


\bibitem[Blog(2018)]%
        {londoncalling2018}
\bibfield{author}{\bibinfo{person}{London~Calling Blog}.}
  \bibinfo{year}{2018}\natexlab{}.
\newblock \bibinfo{title}{New Shok-1 Street Art ‘Stolen Heart’ in Brick
  Lane}.
\newblock
  \bibinfo{howpublished}{\url{https://londoncallingblog.net/2018/06/20/new-shok-1-street-art-stolen-heart-in-brick-lane/}}.
\newblock


\bibitem[Bostick et~al\mbox{.}(2018)]%
        {bostick2018pain}
\bibfield{author}{\bibinfo{person}{Geoff~P Bostick}, \bibinfo{person}{Bruce~D
  Dick}, \bibinfo{person}{Mary Wood}, \bibinfo{person}{Barbara Luckhurst},
  \bibinfo{person}{Julie Tschofen}, {and} \bibinfo{person}{Timothy~W Wideman}.}
  \bibinfo{year}{2018}\natexlab{}.
\newblock \showarticletitle{Pain assessment recommendations for women, made by
  women: a mixed methods study}.
\newblock \bibinfo{journal}{\emph{Pain Medicine}} \bibinfo{volume}{19},
  \bibinfo{number}{6} (\bibinfo{year}{2018}), \bibinfo{pages}{1147--1155}.
\newblock


\bibitem[Burns(2007)]%
        {burns2007gunther}
\bibfield{author}{\bibinfo{person}{Lawrence Burns}.}
  \bibinfo{year}{2007}\natexlab{}.
\newblock \showarticletitle{Gunther von Hagens' BODY WORLDS: Selling beautiful
  education}.
\newblock \bibinfo{journal}{\emph{The American Journal of Bioethics}}
  \bibinfo{volume}{7}, \bibinfo{number}{4} (\bibinfo{year}{2007}),
  \bibinfo{pages}{12--23}.
\newblock


\bibitem[Campbell and Edwards(2012)]%
        {campbell2012ethnic}
\bibfield{author}{\bibinfo{person}{Claudia~M Campbell} {and}
  \bibinfo{person}{Robert~R Edwards}.} \bibinfo{year}{2012}\natexlab{}.
\newblock \showarticletitle{Ethnic differences in pain and pain management}.
\newblock \bibinfo{journal}{\emph{Pain management}} \bibinfo{volume}{2},
  \bibinfo{number}{3} (\bibinfo{year}{2012}), \bibinfo{pages}{219--230}.
\newblock


\bibitem[Chheang et~al\mbox{.}(2023)]%
        {chheang2023towards}
\bibfield{author}{\bibinfo{person}{Vuthea Chheang}, \bibinfo{person}{Rommy
  Marquez-Hernandez}, \bibinfo{person}{Megha Patel}, \bibinfo{person}{Danush
  Rajasekaran}, \bibinfo{person}{Shayla Sharmin}, \bibinfo{person}{Gavin
  Caulfield}, \bibinfo{person}{Behdokht Kiafar}, \bibinfo{person}{Jicheng Li},
  {and} \bibinfo{person}{Roghayeh~Leila Barmaki}.}
  \bibinfo{year}{2023}\natexlab{}.
\newblock \showarticletitle{Towards anatomy education with generative AI-based
  virtual assistants in immersive virtual reality environments}.
\newblock \bibinfo{journal}{\emph{arXiv preprint arXiv:2306.17278}}
  (\bibinfo{year}{2023}).
\newblock


\bibitem[Cipriano and Gleicher(2008)]%
        {cipriano2008text}
\bibfield{author}{\bibinfo{person}{Gregory Cipriano} {and}
  \bibinfo{person}{Michael Gleicher}.} \bibinfo{year}{2008}\natexlab{}.
\newblock \showarticletitle{Text scaffolds for effective surface labeling}.
\newblock \bibinfo{journal}{\emph{IEEE Transactions on Visualization and
  Computer Graphics}} \bibinfo{volume}{14}, \bibinfo{number}{6}
  (\bibinfo{year}{2008}), \bibinfo{pages}{1675--1682}.
\newblock


\bibitem[Collier(2022)]%
        {Collier2022Review}
\bibfield{author}{\bibinfo{person}{Zakiya Collier}.}
  \bibinfo{year}{2022}\natexlab{}.
\newblock \showarticletitle{Review: The {Library} of {Missing} {Datasets}}.
\newblock \bibinfo{journal}{\emph{Reviews in Digital Humanities}}
  \bibinfo{volume}{III}, \bibinfo{number}{4} (\bibinfo{date}{apr 19}
  \bibinfo{year}{2022}).
\newblock
\newblock
\shownote{https://reviewsindh.pubpub.org/pub/library-of-missing-datasets}.


\bibitem[Correll(2019)]%
        {correll2019ethical}
\bibfield{author}{\bibinfo{person}{Michael Correll}.}
  \bibinfo{year}{2019}\natexlab{}.
\newblock \showarticletitle{Ethical dimensions of visualization research}. In
  \bibinfo{booktitle}{\emph{Proceedings of the 2019 CHI conference on human
  factors in computing systems}}. \bibinfo{pages}{1--13}.
\newblock


\bibitem[Dachille-Hey(2012)]%
        {dachille2012case}
\bibfield{author}{\bibinfo{person}{Rae~Erin Dachille-Hey}.}
  \bibinfo{year}{2012}\natexlab{}.
\newblock \showarticletitle{The Case of the Disappearing Blue Women:
  Understanding how Meaning is made in Desi Sangye Gyatso’s Blue Beryl
  paintings}.
\newblock \bibinfo{journal}{\emph{Asian Medicine}} \bibinfo{volume}{6},
  \bibinfo{number}{2} (\bibinfo{year}{2012}), \bibinfo{pages}{293--320}.
\newblock


\bibitem[Daston and Galison(1992)]%
        {daston1992image}
\bibfield{author}{\bibinfo{person}{Lorraine Daston} {and}
  \bibinfo{person}{Peter Galison}.} \bibinfo{year}{1992}\natexlab{}.
\newblock \showarticletitle{The image of objectivity}.
\newblock \bibinfo{journal}{\emph{Representations}} \bibinfo{number}{40}
  (\bibinfo{year}{1992}), \bibinfo{pages}{81--128}.
\newblock


\bibitem[Davidson(2007)]%
        {davidson2007resurrection}
\bibfield{author}{\bibinfo{person}{James~M Davidson}.}
  \bibinfo{year}{2007}\natexlab{}.
\newblock \showarticletitle{“Resurrection men” in Dallas: The illegal use
  of black bodies as medical cadavers (1900--1907)}.
\newblock \bibinfo{journal}{\emph{International Journal of Historical
  Archaeology}}  \bibinfo{volume}{11} (\bibinfo{year}{2007}),
  \bibinfo{pages}{193--220}.
\newblock


\bibitem[D'ignazio and Klein(2020)]%
        {d2020data}
\bibfield{author}{\bibinfo{person}{Catherine D'ignazio} {and}
  \bibinfo{person}{Lauren~F Klein}.} \bibinfo{year}{2020}\natexlab{}.
\newblock \bibinfo{booktitle}{\emph{Data feminism}}.
\newblock \bibinfo{publisher}{MIT press}.
\newblock


\bibitem[Drucker(2011)]%
        {drucker2011humanities}
\bibfield{author}{\bibinfo{person}{Johanna Drucker}.}
  \bibinfo{year}{2011}\natexlab{}.
\newblock \showarticletitle{Humanities approaches to graphical display}.
\newblock \bibinfo{journal}{\emph{Digital Humanities Quarterly}}
  \bibinfo{volume}{5}, \bibinfo{number}{1} (\bibinfo{year}{2011}).
\newblock


\bibitem[Ebenstein et~al\mbox{.}(2016)]%
        {ebenstein2016anatomical}
\bibfield{author}{\bibinfo{person}{Joanna Ebenstein} {et~al\mbox{.}}}
  \bibinfo{year}{2016}\natexlab{}.
\newblock \bibinfo{booktitle}{\emph{The Anatomical Venus}}.
\newblock \bibinfo{publisher}{Thames \& Hudson}.
\newblock


\bibitem[Evergreen(2019)]%
        {evergreen2019}
\bibfield{author}{\bibinfo{person}{Stephanie Evergreen}.}
  \bibinfo{year}{2019}\natexlab{}.
\newblock \bibinfo{title}{Beyond Nightingale: Being a Woman in Data
  Visualization}.
\newblock
  \bibinfo{howpublished}{\url{https://medium.com/nightingale/beyond-nightingale-being-a-woman-in-data-visualization-d7968d171ccf}}.
\newblock


\bibitem[Eysenck(1964)]%
        {eysenck1964principles}
\bibfield{author}{\bibinfo{person}{Hans~J Eysenck}.}
  \bibinfo{year}{1964}\natexlab{}.
\newblock \showarticletitle{Principles and methods of personality description,
  classification and diagnosis}.
\newblock \bibinfo{journal}{\emph{British journal of psychology}}
  \bibinfo{volume}{55}, \bibinfo{number}{3} (\bibinfo{year}{1964}),
  \bibinfo{pages}{284}.
\newblock


\bibitem[Faraone(2011)]%
        {faraone2011magical}
\bibfield{author}{\bibinfo{person}{Christopher~A Faraone}.}
  \bibinfo{year}{2011}\natexlab{}.
\newblock \showarticletitle{Magical and medical approaches to the wandering
  womb in the ancient Greek world}.
\newblock \bibinfo{journal}{\emph{Classical Antiquity}} \bibinfo{volume}{30},
  \bibinfo{number}{1} (\bibinfo{year}{2011}), \bibinfo{pages}{1--32}.
\newblock


\bibitem[Fiorentini(2011)]%
        {fiorentini2011inducing}
\bibfield{author}{\bibinfo{person}{Erna Fiorentini}.}
  \bibinfo{year}{2011}\natexlab{}.
\newblock \showarticletitle{Inducing visibilities: An attempt at Santiago
  Ram{\'o}n y Cajal’s aesthetic epistemology}.
\newblock \bibinfo{journal}{\emph{Studies in History and Philosophy of Science
  Part C: Studies in History and Philosophy of Biological and Biomedical
  Sciences}} \bibinfo{volume}{42}, \bibinfo{number}{4} (\bibinfo{year}{2011}),
  \bibinfo{pages}{391--394}.
\newblock
\urldef\tempurl%
\url{https://doi.org/10.1016/j.shpsc.2011.07.008}
\showDOI{\tempurl}


\bibitem[Fisk(2009)]%
        {fisk2009words}
\bibfield{author}{\bibinfo{person}{Sarah~Anne Fisk}.}
  \bibinfo{year}{2009}\natexlab{}.
\newblock \emph{\bibinfo{title}{When Words Take Lives: The Role of Language in
  the Dehumanization and Devastation of Jews in the Holocaust}}.
\newblock \bibinfo{thesistype}{Ph.\,D. Dissertation}.
  \bibinfo{school}{University of Canterbury. School of Humanities}.
\newblock


\bibitem[Fortuna(2020)]%
        {fortuna2020pseudo}
\bibfield{author}{\bibinfo{person}{Stefania Fortuna}.}
  \bibinfo{year}{2020}\natexlab{}.
\newblock \showarticletitle{Pseudo-Galenic Texts in the Editions of Galen
  (1490-1689)}.
\newblock \bibinfo{journal}{\emph{Medicina nei secoli: Journal of history of
  medicine and medical humanities}} \bibinfo{volume}{32}, \bibinfo{number}{1}
  (\bibinfo{year}{2020}), \bibinfo{pages}{117--138}.
\newblock


\bibitem[Friendly(2006)]%
        {Friendly06hbook}
\bibfield{author}{\bibinfo{person}{Michael Friendly}.}
  \bibinfo{year}{2006}\natexlab{}.
\newblock \showarticletitle{A Brief History of Data Visualization}.
\newblock In \bibinfo{booktitle}{\emph{Handbook of Computational Statistics:
  Data Visualization}}, \bibfield{editor}{\bibinfo{person}{C.~Chen},
  \bibinfo{person}{W.~H\"ardle}, {and} \bibinfo{person}{A~Unwin}} (Eds.).
  Vol.~\bibinfo{volume}{III}. \bibinfo{publisher}{Springer-Verlag},
  \bibinfo{address}{Heidelberg}, \bibinfo{pages}{15--56}.
\newblock


\bibitem[Friendly and Denis(2001)]%
        {friendly2001milestones}
\bibfield{author}{\bibinfo{person}{Michael Friendly} {and}
  \bibinfo{person}{Daniel~J Denis}.} \bibinfo{year}{2001}\natexlab{}.
\newblock \showarticletitle{Milestones in the history of thematic cartography,
  statistical graphics, and data visualization}.
\newblock \bibinfo{journal}{\emph{URL http://www. datavis. ca/milestones}}
  \bibinfo{volume}{32} (\bibinfo{year}{2001}), \bibinfo{pages}{13}.
\newblock


\bibitem[Friendly and Wainer(2021)]%
        {friendly2021history}
\bibfield{author}{\bibinfo{person}{Michael Friendly} {and}
  \bibinfo{person}{Howard Wainer}.} \bibinfo{year}{2021}\natexlab{}.
\newblock \bibinfo{booktitle}{\emph{A history of data visualization and graphic
  communication}}. Vol.~\bibinfo{volume}{56}.
\newblock \bibinfo{publisher}{Harvard University Press Cambridge, MA, EE. UU.}
\newblock


\bibitem[Garrison et~al\mbox{.}(2021)]%
        {garrison2021bioviscom}
\bibfield{author}{\bibinfo{person}{Laura Garrison}, \bibinfo{person}{Monique
  Meuschke}, \bibinfo{person}{Jennifer Fairman}, \bibinfo{person}{Noeska Smit},
  \bibinfo{person}{Bernhard Preim}, {and} \bibinfo{person}{Stefan Bruckner}.}
  \bibinfo{year}{2021}\natexlab{}.
\newblock \showarticletitle{An Exploration of Practice and Preferences for the
  Visual Communication of Biomedical Processes}. In
  \bibinfo{booktitle}{\emph{Proceedings of VCBM}}. \bibinfo{publisher}{The
  Eurographics Association}.
\newblock
\urldef\tempurl%
\url{https://doi.org/10.2312/vcbm.20211339}
\showDOI{\tempurl}


\bibitem[Ghosh(2015a)]%
        {ghosh2015evolution}
\bibfield{author}{\bibinfo{person}{Sanjib~Kumar Ghosh}.}
  \bibinfo{year}{2015}\natexlab{a}.
\newblock \showarticletitle{Evolution of illustrations in anatomy: A study from
  the classical period in E urope to modern times}.
\newblock \bibinfo{journal}{\emph{Anatomical Sciences Education}}
  \bibinfo{volume}{8}, \bibinfo{number}{2} (\bibinfo{year}{2015}),
  \bibinfo{pages}{175--188}.
\newblock


\bibitem[Ghosh(2015b)]%
        {ghosh2015human}
\bibfield{author}{\bibinfo{person}{Sanjib~Kumar Ghosh}.}
  \bibinfo{year}{2015}\natexlab{b}.
\newblock \showarticletitle{Human cadaveric dissection: a historical account
  from ancient Greece to the modern era}.
\newblock \bibinfo{journal}{\emph{Anatomy \& Cell Biology}}
  \bibinfo{volume}{48}, \bibinfo{number}{3} (\bibinfo{year}{2015}),
  \bibinfo{pages}{153--169}.
\newblock
\urldef\tempurl%
\url{https://doi.org/10.5115/acb.2015.48.3.153}
\showDOI{\tempurl}


\bibitem[Ghosh(2017)]%
        {ghosh2017cadaveric}
\bibfield{author}{\bibinfo{person}{Sanjib~Kumar Ghosh}.}
  \bibinfo{year}{2017}\natexlab{}.
\newblock \showarticletitle{Cadaveric dissection as an educational tool for
  anatomical sciences in the 21st century}.
\newblock \bibinfo{journal}{\emph{Anatomical sciences education}}
  \bibinfo{volume}{10}, \bibinfo{number}{3} (\bibinfo{year}{2017}),
  \bibinfo{pages}{286--299}.
\newblock
\urldef\tempurl%
\url{https://doi.org/10.1002/ase.1649}
\showDOI{\tempurl}


\bibitem[Green(2010)]%
        {green2010working}
\bibfield{author}{\bibinfo{person}{Alexa Green}.}
  \bibinfo{year}{2010}\natexlab{}.
\newblock \showarticletitle{Working ethics: William Beaumont, Alexis St.
  Martin, and medical research in Antebellum America}.
\newblock \bibinfo{journal}{\emph{Bulletin of the History of Medicine}}
  (\bibinfo{year}{2010}), \bibinfo{pages}{193--216}.
\newblock
\urldef\tempurl%
\url{https://doi.org/10.1353/bhm.0.0341}
\showDOI{\tempurl}


\bibitem[Green(2000)]%
        {green2000diseases}
\bibfield{author}{\bibinfo{person}{Monica~Helen Green}.}
  \bibinfo{year}{2000}\natexlab{}.
\newblock \showarticletitle{From" Diseases of Women" to" Secrets of Women": The
  Transformation of Gynecological Literature in the Later Middle Ages}.
\newblock \bibinfo{journal}{\emph{Journal of Medieval and Early Modern
  Studies}} \bibinfo{volume}{30}, \bibinfo{number}{1} (\bibinfo{year}{2000}),
  \bibinfo{pages}{5--39}.
\newblock


\bibitem[Habicht et~al\mbox{.}(2018)]%
        {habicht2018bodies}
\bibfield{author}{\bibinfo{person}{Juri~L Habicht}, \bibinfo{person}{Claudia
  Kiessling}, {and} \bibinfo{person}{Andreas Winkelmann}.}
  \bibinfo{year}{2018}\natexlab{}.
\newblock \showarticletitle{Bodies for anatomy education in medical schools: an
  overview of the sources of cadavers worldwide}.
\newblock \bibinfo{journal}{\emph{Academic Medicine}} \bibinfo{volume}{93},
  \bibinfo{number}{9} (\bibinfo{year}{2018}), \bibinfo{pages}{1293}.
\newblock
\urldef\tempurl%
\url{https://doi.org/10.1097/ACM.0000000000002227}
\showDOI{\tempurl}


\bibitem[Halperin(2007)]%
        {halperin2007poor}
\bibfield{author}{\bibinfo{person}{Edward~C Halperin}.}
  \bibinfo{year}{2007}\natexlab{}.
\newblock \showarticletitle{The poor, the black, and the marginalized as the
  source of cadavers in United States anatomical education}.
\newblock \bibinfo{journal}{\emph{Clinical Anatomy: The Official Journal of the
  American Association of Clinical Anatomists and the British Association of
  Clinical Anatomists}} \bibinfo{volume}{20}, \bibinfo{number}{5}
  (\bibinfo{year}{2007}), \bibinfo{pages}{489--495}.
\newblock


\bibitem[Halperin(2009)]%
        {halperin2009pornographic}
\bibfield{author}{\bibinfo{person}{Edward~C Halperin}.}
  \bibinfo{year}{2009}\natexlab{}.
\newblock \showarticletitle{The pornographic anatomy book? The curious tale of
  the Anatomical Basis of Medical Practice}.
\newblock \bibinfo{journal}{\emph{Academic Medicine}} \bibinfo{volume}{84},
  \bibinfo{number}{2} (\bibinfo{year}{2009}), \bibinfo{pages}{278--283}.
\newblock


\bibitem[Halpern(2015)]%
        {halpern2015beautiful}
\bibfield{author}{\bibinfo{person}{Orit Halpern}.}
  \bibinfo{year}{2015}\natexlab{}.
\newblock \bibinfo{booktitle}{\emph{Beautiful data: A history of vision and
  reason since 1945}}.
\newblock \bibinfo{publisher}{Duke University Press}.
\newblock


\bibitem[Haraway(1988)]%
        {haraway1988situated}
\bibfield{author}{\bibinfo{person}{Donna Haraway}.}
  \bibinfo{year}{1988}\natexlab{}.
\newblock \showarticletitle{Situated knowledges: The science question in
  feminism and the privilege of partial perspective}.
\newblock \bibinfo{journal}{\emph{Feminist Studies}} \bibinfo{volume}{14},
  \bibinfo{number}{3} (\bibinfo{year}{1988}), \bibinfo{pages}{575--599}.
\newblock


\bibitem[Hartnell(2016)]%
        {hartnell_many_nodate}
\bibfield{author}{\bibinfo{person}{Jack Hartnell}.}
  \bibinfo{year}{2016}\natexlab{}.
\newblock \bibinfo{title}{The {Many} {Lives} of the {Medieval} {Wound} {Man}}.
\newblock
\newblock
\urldef\tempurl%
\url{https://publicdomainreview.org/essay/the-many-lives-of-the-medieval-wound-man/}
\showURL{%
\tempurl}


\bibitem[Hartnell(2017)]%
        {hartnell2017wording}
\bibfield{author}{\bibinfo{person}{Jack Hartnell}.}
  \bibinfo{year}{2017}\natexlab{}.
\newblock \showarticletitle{Wording the Wound Man}.
\newblock \bibinfo{journal}{\emph{British Art Studies}} \bibinfo{number}{6}
  (\bibinfo{year}{2017}).
\newblock


\bibitem[Hoffman et~al\mbox{.}(2016)]%
        {hoffman2016racial}
\bibfield{author}{\bibinfo{person}{Kelly~M Hoffman}, \bibinfo{person}{Sophie
  Trawalter}, \bibinfo{person}{Jordan~R Axt}, {and} \bibinfo{person}{M~Norman
  Oliver}.} \bibinfo{year}{2016}\natexlab{}.
\newblock \showarticletitle{Racial bias in pain assessment and treatment
  recommendations, and false beliefs about biological differences between
  blacks and whites}.
\newblock \bibinfo{journal}{\emph{Proceedings of the National Academy of
  Sciences}} \bibinfo{volume}{113}, \bibinfo{number}{16}
  (\bibinfo{year}{2016}), \bibinfo{pages}{4296--4301}.
\newblock


\bibitem[Howell(2016)]%
        {howell2016early}
\bibfield{author}{\bibinfo{person}{Joel~D Howell}.}
  \bibinfo{year}{2016}\natexlab{}.
\newblock \showarticletitle{Early clinical use of the X-ray}.
\newblock \bibinfo{journal}{\emph{Transactions of the American Clinical and
  Climatological Association}}  \bibinfo{volume}{127} (\bibinfo{year}{2016}),
  \bibinfo{pages}{341}.
\newblock


\bibitem[Hutton(2015)]%
        {hutton2015study}
\bibfield{author}{\bibinfo{person}{Fiona Hutton}.}
  \bibinfo{year}{2015}\natexlab{}.
\newblock \bibinfo{booktitle}{\emph{The Study of Anatomy in Britain,
  1700--1900}}.
\newblock \bibinfo{publisher}{Routledge}.
\newblock


\bibitem[Jack(2009)]%
        {jack2009pedagogy}
\bibfield{author}{\bibinfo{person}{Jordynn Jack}.}
  \bibinfo{year}{2009}\natexlab{}.
\newblock \showarticletitle{A pedagogy of sight: Microscopic vision in Robert
  Hooke's Micrographia}.
\newblock \bibinfo{journal}{\emph{Quarterly Journal of Speech}}
  \bibinfo{volume}{95}, \bibinfo{number}{2} (\bibinfo{year}{2009}),
  \bibinfo{pages}{192--209}.
\newblock
\urldef\tempurl%
\url{https://doi.org/10.1080/00335630902842079}
\showDOI{\tempurl}


\bibitem[J{\"a}nicke et~al\mbox{.}(2015)]%
        {janicke2015close}
\bibfield{author}{\bibinfo{person}{Stefan J{\"a}nicke}, \bibinfo{person}{Greta
  Franzini}, \bibinfo{person}{Muhammad~Faisal Cheema}, {and}
  \bibinfo{person}{Gerik Scheuermann}.} \bibinfo{year}{2015}\natexlab{}.
\newblock \showarticletitle{On Close and Distant Reading in Digital Humanities:
  A Survey and Future Challenges.}
\newblock \bibinfo{journal}{\emph{EuroVis (STARs)}}  \bibinfo{volume}{2015}
  (\bibinfo{year}{2015}), \bibinfo{pages}{83--103}.
\newblock


\bibitem[Kearns and Kearns(2020)]%
        {kearns2020role}
\bibfield{author}{\bibinfo{person}{Cil{\'e}in Kearns} {and}
  \bibinfo{person}{Nethmi Kearns}.} \bibinfo{year}{2020}\natexlab{}.
\newblock \showarticletitle{The role of comics in public health communication
  during the COVID-19 pandemic}.
\newblock \bibinfo{journal}{\emph{Journal of visual communication in medicine}}
  \bibinfo{volume}{43}, \bibinfo{number}{3} (\bibinfo{year}{2020}),
  \bibinfo{pages}{139--149}.
\newblock


\bibitem[Kemp(1970)]%
        {kemp1970drawing}
\bibfield{author}{\bibinfo{person}{Martin Kemp}.}
  \bibinfo{year}{1970}\natexlab{}.
\newblock \showarticletitle{A drawing for the Fabrica; and some thoughts upon
  the Vesalius muscle-men}.
\newblock \bibinfo{journal}{\emph{Medical history}} \bibinfo{volume}{14},
  \bibinfo{number}{3} (\bibinfo{year}{1970}), \bibinfo{pages}{277--288}.
\newblock


\bibitem[Kikkeri and Nagalli(2022)]%
        {kikkeri2022migraine}
\bibfield{author}{\bibinfo{person}{Nidhi~Shankar Kikkeri} {and}
  \bibinfo{person}{Shivaraj Nagalli}.} \bibinfo{year}{2022}\natexlab{}.
\newblock \showarticletitle{Migraine with aura}.
\newblock In \bibinfo{booktitle}{\emph{StatPearls [Internet]}}.
  \bibinfo{publisher}{StatPearls Publishing}.
\newblock


\bibitem[Klein(2022)]%
        {klein2022data}
\bibfield{author}{\bibinfo{person}{Lauren Klein}.}
  \bibinfo{year}{2022}\natexlab{}.
\newblock \showarticletitle{What Data Visualization Reveals: Elizabeth Palmer
  Peabody and the Work of Knowledge Production}.
\newblock \bibinfo{journal}{\emph{Harvard Data Science Review}}
  \bibinfo{volume}{4}, \bibinfo{number}{2} (\bibinfo{year}{2022}),
  \bibinfo{pages}{1--34}.
\newblock


\bibitem[Kuriyama and Kuriyama(1999)]%
        {kuriyama1999expressiveness}
\bibfield{author}{\bibinfo{person}{Shigehisa Kuriyama} {and}
  \bibinfo{person}{Shigehisa Kuriyama}.} \bibinfo{year}{1999}\natexlab{}.
\newblock \showarticletitle{The expressiveness of the body and the divergence
  of Greek and Chinese medicine}.
\newblock  (\bibinfo{year}{1999}).
\newblock


\bibitem[Lanska(2015)]%
        {lanska2015evolution}
\bibfield{author}{\bibinfo{person}{D Lanska}.} \bibinfo{year}{2015}\natexlab{}.
\newblock \showarticletitle{The evolution of Vesalius’s perspective on
  Galen’s anatomy}.
\newblock \bibinfo{journal}{\emph{Hist Med}} \bibinfo{volume}{2},
  \bibinfo{number}{1} (\bibinfo{year}{2015}), \bibinfo{pages}{13--26}.
\newblock


\bibitem[Lee et~al\mbox{.}(2021)]%
        {lee2021viral}
\bibfield{author}{\bibinfo{person}{Crystal Lee}, \bibinfo{person}{Tanya Yang},
  \bibinfo{person}{Gabrielle~D Inchoco}, \bibinfo{person}{Graham~M Jones},
  {and} \bibinfo{person}{Arvind Satyanarayan}.}
  \bibinfo{year}{2021}\natexlab{}.
\newblock \showarticletitle{Viral visualizations: How coronavirus skeptics use
  orthodox data practices to promote unorthodox science online}. In
  \bibinfo{booktitle}{\emph{Proceedings of the 2021 CHI conference on human
  factors in computing systems}}. \bibinfo{pages}{1--18}.
\newblock


\bibitem[Limm(2022)]%
        {viralbipocanatomy}
\bibfield{author}{\bibinfo{person}{David Limm}.}
  \bibinfo{year}{2022}\natexlab{}.
\newblock \bibinfo{title}{The Creator of a Viral Black Fetus Medical
  Illustration Blends Art and Activism}.
\newblock
  \bibinfo{howpublished}{\url{https://healthcity.bmc.org/policy-and-industry/creator-viral-black-fetus-medical-illustration-blends-art-and-activism}}.
\newblock


\bibitem[Luesink(2017)]%
        {luesink2017anatomy}
\bibfield{author}{\bibinfo{person}{David Luesink}.}
  \bibinfo{year}{2017}\natexlab{}.
\newblock \showarticletitle{Anatomy and the reconfiguration of life and death
  in Republican China}.
\newblock \bibinfo{journal}{\emph{The Journal of Asian Studies}}
  \bibinfo{volume}{76}, \bibinfo{number}{4} (\bibinfo{year}{2017}),
  \bibinfo{pages}{1009--1034}.
\newblock
\urldef\tempurl%
\url{https://doi.org/10.1163/9789004397620_021}
\showDOI{\tempurl}


\bibitem[Lupi and King(2018)]%
        {lupi2018bruises}
\bibfield{author}{\bibinfo{person}{Giorgia Lupi} {and} \bibinfo{person}{Kaki
  King}.} \bibinfo{year}{2018}\natexlab{}.
\newblock \showarticletitle{Bruises: The data we don’t see}.
\newblock \bibinfo{journal}{\emph{Hematomas: Los Datos que No Vemos] Medium:
  Neuroscience}}  \bibinfo{volume}{31} (\bibinfo{year}{2018}).
\newblock


\bibitem[Marchese(2012)]%
        {marchese2012origins}
\bibfield{author}{\bibinfo{person}{Francis~T Marchese}.}
  \bibinfo{year}{2012}\natexlab{}.
\newblock \showarticletitle{The origins and rise of medieval information
  visualization}. In \bibinfo{booktitle}{\emph{2012 16th International
  Conference on Information Visualisation}}. IEEE, \bibinfo{pages}{389--395}.
\newblock


\bibitem[Marchese(2013)]%
        {marchese2013medieval}
\bibfield{author}{\bibinfo{person}{Francis~T Marchese}.}
  \bibinfo{year}{2013}\natexlab{}.
\newblock \showarticletitle{Medieval information visualization}.
\newblock \bibinfo{journal}{\emph{Proceedings of the IEEE VIS Arts Program
  (VISAP)}} (\bibinfo{year}{2013}).
\newblock


\bibitem[Marx(1920)]%
        {marx1920poverty}
\bibfield{author}{\bibinfo{person}{Karl Marx}.}
  \bibinfo{year}{1920}\natexlab{}.
\newblock \bibinfo{booktitle}{\emph{The Poverty of Philosophy}}.
\newblock \bibinfo{publisher}{CH Kerr}.
\newblock


\bibitem[Marx(1926)]%
        {marx1926eighteenth}
\bibfield{author}{\bibinfo{person}{Karl Marx}.}
  \bibinfo{year}{1926}\natexlab{}.
\newblock \bibinfo{booktitle}{\emph{The eighteenth brumaire of Louis
  Bonaparte}}.
\newblock \bibinfo{publisher}{International Publishers}.
\newblock


\bibitem[Matuk(2006)]%
        {matuk2006seeing}
\bibfield{author}{\bibinfo{person}{Camillia Matuk}.}
  \bibinfo{year}{2006}\natexlab{}.
\newblock \showarticletitle{Seeing the body: The divergence of ancient Chinese
  and western medical illustration}.
\newblock \bibinfo{journal}{\emph{Journal of Biocommunication}}
  \bibinfo{volume}{32}, \bibinfo{number}{1} (\bibinfo{year}{2006}),
  \bibinfo{pages}{1--8}.
\newblock


\bibitem[Mbemb{\'e} and Meintjes(2003)]%
        {mbembe2003necropolitics}
\bibfield{author}{\bibinfo{person}{J-A Mbemb{\'e}} {and} \bibinfo{person}{Libby
  Meintjes}.} \bibinfo{year}{2003}\natexlab{}.
\newblock \showarticletitle{Necropolitics}.
\newblock \bibinfo{journal}{\emph{Public culture}} \bibinfo{volume}{15},
  \bibinfo{number}{1} (\bibinfo{year}{2003}), \bibinfo{pages}{11--40}.
\newblock


\bibitem[McClusky~III et~al\mbox{.}(1997)]%
        {mcclusky1997hepatic}
\bibfield{author}{\bibinfo{person}{David~A McClusky~III},
  \bibinfo{person}{Lee~J Skandalakis}, \bibinfo{person}{Gene~L Colborn}, {and}
  \bibinfo{person}{John~E Skandalakis}.} \bibinfo{year}{1997}\natexlab{}.
\newblock \showarticletitle{Hepatic surgery and hepatic surgical anatomy:
  historical partners in progress}.
\newblock \bibinfo{journal}{\emph{World journal of surgery}}
  \bibinfo{volume}{21} (\bibinfo{year}{1997}), \bibinfo{pages}{330--342}.
\newblock


\bibitem[Meyer and Dykes(2019)]%
        {meyer2019criteria}
\bibfield{author}{\bibinfo{person}{Miriah Meyer} {and} \bibinfo{person}{Jason
  Dykes}.} \bibinfo{year}{2019}\natexlab{}.
\newblock \showarticletitle{Criteria for rigor in visualization design study}.
\newblock \bibinfo{journal}{\emph{IEEE transactions on visualization and
  computer graphics}} \bibinfo{volume}{26}, \bibinfo{number}{1}
  (\bibinfo{year}{2019}), \bibinfo{pages}{87--97}.
\newblock


\bibitem[Morais et~al\mbox{.}(2020)]%
        {morais2020showing}
\bibfield{author}{\bibinfo{person}{Luiz Morais}, \bibinfo{person}{Yvonne
  Jansen}, \bibinfo{person}{Nazareno Andrade}, {and} \bibinfo{person}{Pierre
  Dragicevic}.} \bibinfo{year}{2020}\natexlab{}.
\newblock \showarticletitle{Showing data about people: A design space of
  anthropographics}.
\newblock \bibinfo{journal}{\emph{IEEE Transactions on Visualization and
  Computer Graphics}} \bibinfo{volume}{28}, \bibinfo{number}{3}
  (\bibinfo{year}{2020}), \bibinfo{pages}{1661--1679}.
\newblock


\bibitem[Munson(1996)]%
        {munson1996note}
\bibfield{author}{\bibinfo{person}{David~C Munson}.}
  \bibinfo{year}{1996}\natexlab{}.
\newblock \showarticletitle{A note on Lena}.
\newblock \bibinfo{journal}{\emph{IEEE Transactions on Image Processing}}
  \bibinfo{volume}{5}, \bibinfo{number}{1} (\bibinfo{year}{1996}),
  \bibinfo{pages}{3--3}.
\newblock


\bibitem[Netter(2022)]%
        {netter2022netter}
\bibfield{author}{\bibinfo{person}{Frank~H Netter}.}
  \bibinfo{year}{2022}\natexlab{}.
\newblock \bibinfo{booktitle}{\emph{Netter Atlas of Human Anatomy: A Systems
  Approach-E-Book: paperback+ eBook}}.
\newblock \bibinfo{publisher}{Elsevier Health Sciences}.
\newblock


\bibitem[of~Medical~Illustrators(2023)]%
        {amidiversityfellow}
\bibfield{author}{\bibinfo{person}{Association of Medical~Illustrators}.}
  \bibinfo{year}{2023}\natexlab{}.
\newblock \bibinfo{title}{AMI Diversity Scholarship}.
\newblock \bibinfo{howpublished}{\url{https://awards.ami.org/}}.
\newblock


\bibitem[of~Toronto Thomas Fisher Rare Book~Library(2023)]%
        {ut2023emergingpatterns}
\bibfield{author}{\bibinfo{person}{University of Toronto Thomas Fisher Rare
  Book~Library}.} \bibinfo{year}{2023}\natexlab{}.
\newblock \bibinfo{title}{Emerging Patterns: Data Visualization Throughout
  History}.
\newblock
  \bibinfo{howpublished}{\url{https://fisher.library.utoronto.ca/exhibition/emerging-patterns-data-visualization-throughout-history}}.
\newblock


\bibitem[on~the Web(2016)]%
        {nlmhua}
\bibfield{author}{\bibinfo{person}{NLM Historical~Anatomies on~the Web}.}
  \bibinfo{year}{2016}\natexlab{}.
\newblock \bibinfo{title}{Shou, Hua. Jushikei hakki (Shi si jing fa hui.
  Japanese \& Chinese)}.
\newblock
  \bibinfo{howpublished}{\url{https://www.nlm.nih.gov/exhibition/historicalanatomies/hua_bio.html}}.
\newblock


\bibitem[Onuoha(2016)]%
        {onuoha2016library}
\bibfield{author}{\bibinfo{person}{Mimi Onuoha}.}
  \bibinfo{year}{2016}\natexlab{}.
\newblock \showarticletitle{The library of missing datasets}.
\newblock \bibinfo{journal}{\emph{Library Catalog: mimionuoha. com}}
  (\bibinfo{year}{2016}).
\newblock


\bibitem[{\"O}zveren and A{\u{g}}{\i}r(2015)]%
        {ozveren2015review}
\bibfield{author}{\bibinfo{person}{Ey{\"u}p {\"O}zveren} {and}
  \bibinfo{person}{Seven A{\u{g}}{\i}r}.} \bibinfo{year}{2015}\natexlab{}.
\newblock \bibinfo{title}{A Review of ``A History of Ottoman Economic Thought.
  Developments Before the Nineteenth Century''', by Fatih Ermi{\c{s}}: London
  and New York: Routledge, 2013, 232 pp.,{\pounds} 78.42, ISBN 978-0415540063}.
\newblock
\newblock


\bibitem[Preim and Botha(2013)]%
        {preim2013visual}
\bibfield{author}{\bibinfo{person}{Bernhard Preim} {and}
  \bibinfo{person}{Charl~P Botha}.} \bibinfo{year}{2013}\natexlab{}.
\newblock \bibinfo{booktitle}{\emph{Visual computing for medicine: theory,
  algorithms, and applications}}.
\newblock \bibinfo{publisher}{Newnes}.
\newblock


\bibitem[Ramellini(2013)]%
        {ramellini2013life}
\bibfield{author}{\bibinfo{person}{Pietro Ramellini}.}
  \bibinfo{year}{2013}\natexlab{}.
\newblock \showarticletitle{Life: Science, Philosophy, Theology}.
\newblock \bibinfo{journal}{\emph{Studia Bioethica}} \bibinfo{volume}{6},
  \bibinfo{number}{2} (\bibinfo{year}{2013}).
\newblock


\bibitem[Reichle et~al\mbox{.}(2009)]%
        {reichle2009art}
\bibfield{author}{\bibinfo{person}{Ingeborg Reichle}, \bibinfo{person}{Robert
  Zwijnenberg}, {and} \bibinfo{person}{Gloria Custance}.}
  \bibinfo{year}{2009}\natexlab{}.
\newblock \bibinfo{booktitle}{\emph{Art in the age of technoscience: Genetic
  engineering, robotics, and artificial life in contemporary art}}.
\newblock \bibinfo{publisher}{Springer}.
\newblock


\bibitem[Rendgen(2019)]%
        {rendgen2019historyinfographic}
\bibfield{author}{\bibinfo{person}{Sandra Rendgen}.}
  \bibinfo{year}{2019}\natexlab{}.
\newblock \bibinfo{booktitle}{\emph{History of Information Graphics}}.
\newblock \bibinfo{publisher}{TASCHEN}, \bibinfo{address}{Hohenzollernring 53,
  D–50672 Cologne}.
\newblock


\bibitem[Richards(2003)]%
        {richards2003argument}
\bibfield{author}{\bibinfo{person}{Anne~R Richards}.}
  \bibinfo{year}{2003}\natexlab{}.
\newblock \showarticletitle{Argument and authority in the visual
  representations of science}.
\newblock \bibinfo{journal}{\emph{Technical Communication Quarterly}}
  \bibinfo{volume}{12}, \bibinfo{number}{2} (\bibinfo{year}{2003}),
  \bibinfo{pages}{183--206}.
\newblock


\bibitem[Riggs(1998)]%
        {riggs1998should}
\bibfield{author}{\bibinfo{person}{Garrett Riggs}.}
  \bibinfo{year}{1998}\natexlab{}.
\newblock \showarticletitle{What should we do about Eduard Pernkopf's atlas?}
\newblock \bibinfo{journal}{\emph{Academic Medicine: Journal of the Association
  of American Medical Colleges}} \bibinfo{volume}{73}, \bibinfo{number}{4}
  (\bibinfo{year}{1998}), \bibinfo{pages}{380--386}.
\newblock


\bibitem[Ropinski(2009)]%
        {ropinski09illustravis}
\bibfield{author}{\bibinfo{person}{Timo Ropinski}.}
  \bibinfo{year}{2009}\natexlab{}.
\newblock \showarticletitle{Annotation for presentation: Integrating text in
  medical illustrations} \emph{(\bibinfo{series}{IllustraVis})}.
\newblock


\bibitem[Ropinski et~al\mbox{.}(2007)]%
        {RopinskiPRH07}
\bibfield{author}{\bibinfo{person}{Timo Ropinski},
  \bibinfo{person}{J{\"{o}}rg{-}Stefan Pra{\ss}ni}, \bibinfo{person}{Jan
  Roters}, {and} \bibinfo{person}{Klaus~H. Hinrichs}.}
  \bibinfo{year}{2007}\natexlab{}.
\newblock \showarticletitle{Internal Labels as Shape Cues for Medical
  Illustration}. In \bibinfo{booktitle}{\emph{12th International Fall Workshop
  on Vision, Modeling, and Visualization, {VMV} 2007, Saarbr{\"{u}}cken,
  Germany, November 7-9, 2007}}, \bibfield{editor}{\bibinfo{person}{Hendrik
  P.~A. Lensch}, \bibinfo{person}{Bodo Rosenhahn},
  \bibinfo{person}{Hans{-}Peter Seidel}, \bibinfo{person}{Philipp Slusallek},
  {and} \bibinfo{person}{Joachim Weickert}} (Eds.). \bibinfo{publisher}{Aka
  GmbH}, \bibinfo{pages}{203--212}.
\newblock


\bibitem[Ruiz(2016)]%
        {ruiz2016tedmed}
\bibfield{author}{\bibinfo{person}{Vanessa Ruiz}.}
  \bibinfo{year}{2016}\natexlab{}.
\newblock \bibinfo{booktitle}{\emph{The spellbinding art of human anatomy}}.
\newblock TEDMED.
\newblock
\urldef\tempurl%
\url{https://www.youtube.com/watch?v=M_X0uwAG2Jc&t=679s}
\showURL{%
\tempurl}


\bibitem[Sabernig(2017)]%
        {sabernig2017vulnerable}
\bibfield{author}{\bibinfo{person}{Katharina~A Sabernig}.}
  \bibinfo{year}{2017}\natexlab{}.
\newblock \showarticletitle{Vulnerable Parts: Locating and Defining Vital Areas
  of the Body in Tibetan Medicine}.
\newblock \bibinfo{journal}{\emph{Asian Medicine}} \bibinfo{volume}{12},
  \bibinfo{number}{1-2} (\bibinfo{year}{2017}), \bibinfo{pages}{86--118}.
\newblock


\bibitem[Sahlgrenska~Academy(2022)]%
        {bodyneedsUGot}
\bibfield{author}{\bibinfo{person}{University of~Gothenburg
  Sahlgrenska~Academy}.} \bibinfo{year}{2022}\natexlab{}.
\newblock \bibinfo{title}{Surgeons need more bodies to practice on}.
\newblock
  \bibinfo{howpublished}{\url{https://www.gu.se/en/news/surgeons-need-more-bodies-to-practice-on}}.
\newblock
\newblock
\shownote{Accessed: 2023-12-03}.


\bibitem[Sappol(2004)]%
        {sappol2004morbid}
\bibfield{author}{\bibinfo{person}{Michael Sappol}.}
  \bibinfo{year}{2004}\natexlab{}.
\newblock \showarticletitle{‘Morbid Curiosity’: The Decline and Fall of the
  Popular Anatomical Museum}.
\newblock \bibinfo{journal}{\emph{Common-place}} \bibinfo{volume}{4},
  \bibinfo{number}{2} (\bibinfo{year}{2004}).
\newblock


\bibitem[Sappol and National Library~of Health~(Bethesda(2006)]%
        {sappol2006dream}
\bibfield{author}{\bibinfo{person}{Michael Sappol} {and} \bibinfo{person}{Md)
  National Library~of Health~(Bethesda}.} \bibinfo{year}{2006}\natexlab{}.
\newblock \bibinfo{booktitle}{\emph{Dream anatomy}}.
\newblock \bibinfo{publisher}{US Department of Health and Human Services,
  National Institutes of Health~…}.
\newblock


\bibitem[Shoja and Tubbs(2007)]%
        {shoja2007history}
\bibfield{author}{\bibinfo{person}{Mohammadali~M Shoja} {and}
  \bibinfo{person}{R~Shane Tubbs}.} \bibinfo{year}{2007}\natexlab{}.
\newblock \showarticletitle{The history of anatomy in Persia}.
\newblock \bibinfo{journal}{\emph{Journal of anatomy}} \bibinfo{volume}{210},
  \bibinfo{number}{4} (\bibinfo{year}{2007}), \bibinfo{pages}{359--378}.
\newblock


\bibitem[SHOK-1(2023)]%
        {SHOK2023}
\bibfield{author}{\bibinfo{person}{SHOK-1}.} \bibinfo{year}{2023}\natexlab{}.
\newblock \bibinfo{title}{SHOK-1 Art Retail Biography}.
\newblock \bibinfo{howpublished}{\url{https://store.shok1.com/pages/about-1}}.
\newblock


\bibitem[Singer(1917)]%
        {gutenberg2014historymethods}
\bibfield{author}{\bibinfo{person}{Charles Singer}.}
  \bibinfo{year}{1917}\natexlab{}.
\newblock \bibinfo{title}{Project Gutenberg's Studies in the History and Method
  of Science}.
\newblock
  \bibinfo{howpublished}{\url{https://www.gutenberg.org/files/46572/46572-h/46572-h.htm}}.
\newblock


\bibitem[Singer(1946)]%
        {singer1946some}
\bibfield{author}{\bibinfo{person}{Charles Singer}.}
  \bibinfo{year}{1946}\natexlab{}.
\newblock \showarticletitle{Some Galenic and animal sources of Vesalius}.
\newblock \bibinfo{journal}{\emph{Journal of the History of Medicine and Allied
  Sciences}} \bibinfo{volume}{1}, \bibinfo{number}{1} (\bibinfo{year}{1946}),
  \bibinfo{pages}{6--24}.
\newblock


\bibitem[Singer(1949)]%
        {singer1949galen}
\bibfield{author}{\bibinfo{person}{Charles Singer}.}
  \bibinfo{year}{1949}\natexlab{}.
\newblock \showarticletitle{Galen as a Modern}.
\newblock \bibinfo{journal}{\emph{Proceedings of the Royal Society of
  Medicine}} \bibinfo{volume}{42}, \bibinfo{number}{8} (\bibinfo{year}{1949}),
  \bibinfo{pages}{563--570}.
\newblock


\bibitem[Slobin(2014)]%
        {slobin2014}
\bibfield{author}{\bibinfo{person}{Sarah Slobin}.}
  \bibinfo{year}{2014}\natexlab{}.
\newblock \showarticletitle{What If the Data Visualization is Actually People?}
\newblock \bibinfo{journal}{\emph{Source}} (\bibinfo{date}{Apr}
  \bibinfo{year}{2014}).
\newblock
\urldef\tempurl%
\url{https://source.opennews.org/articles/what-if-data-visualization-actually-people/}
\showURL{%
\tempurl}


\bibitem[Standring(2016)]%
        {standring2016brief}
\bibfield{author}{\bibinfo{person}{Susan Standring}.}
  \bibinfo{year}{2016}\natexlab{}.
\newblock \showarticletitle{A brief history of topographical anatomy}.
\newblock \bibinfo{journal}{\emph{Journal of Anatomy}} \bibinfo{volume}{229},
  \bibinfo{number}{1} (\bibinfo{year}{2016}), \bibinfo{pages}{32--62}.
\newblock


\bibitem[Sullivan(1994)]%
        {sullivan1994sanguine}
\bibfield{author}{\bibinfo{person}{Robert~B Sullivan}.}
  \bibinfo{year}{1994}\natexlab{}.
\newblock \showarticletitle{Sanguine practices: a historical and
  historiographic reconsideration of heroic therapy in the age of Rush}.
\newblock \bibinfo{journal}{\emph{Bulletin of the History of Medicine}}
  \bibinfo{volume}{68}, \bibinfo{number}{2} (\bibinfo{year}{1994}),
  \bibinfo{pages}{211--234}.
\newblock


\bibitem[Van~Alphen et~al\mbox{.}(1995)]%
        {van1995oriental}
\bibfield{author}{\bibinfo{person}{Jan Van~Alphen}, \bibinfo{person}{Anthony
  Aris}, {and} \bibinfo{person}{Mark De~Fraeye}.}
  \bibinfo{year}{1995}\natexlab{}.
\newblock \bibinfo{booktitle}{\emph{Oriental medicine: An illustrated guide to
  the Asian arts of healing}}.
\newblock \bibinfo{publisher}{Serindia Publications, London}.
\newblock


\bibitem[Von~Staden(1975)]%
        {von1975experiment}
\bibfield{author}{\bibinfo{person}{Heinrich Von~Staden}.}
  \bibinfo{year}{1975}\natexlab{}.
\newblock \showarticletitle{Experiment and experience in Hellenistic medicine}.
\newblock \bibinfo{journal}{\emph{Bulletin of the Institute of Classical
  Studies}} \bibinfo{number}{22} (\bibinfo{year}{1975}),
  \bibinfo{pages}{178--199}.
\newblock
\urldef\tempurl%
\url{https://doi.org/10.1111/j.2041-5370.1975.tb00340.x}
\showDOI{\tempurl}


\bibitem[Von~Staden(1992)]%
        {von1992discovery}
\bibfield{author}{\bibinfo{person}{Heinrich Von~Staden}.}
  \bibinfo{year}{1992}\natexlab{}.
\newblock \showarticletitle{The discovery of the body: human dissection and its
  cultural contexts in ancient Greece.}
\newblock \bibinfo{journal}{\emph{The Yale Journal of Biology and Medicine}}
  \bibinfo{volume}{65}, \bibinfo{number}{3} (\bibinfo{year}{1992}),
  \bibinfo{pages}{223}.
\newblock


\bibitem[White et~al\mbox{.}(2022)]%
        {nyt}
\bibfield{author}{\bibinfo{person}{Jeremy White}, \bibinfo{person}{Amy Harmon},
  \bibinfo{person}{Danielle Ivory}, \bibinfo{person}{Lauren Leatherby},
  \bibinfo{person}{Albert Sun}, {and} \bibinfo{person}{Sarah Almukhtar}.}
  \bibinfo{year}{2022}\natexlab{}.
\newblock \showarticletitle{How America Lost One Million People}.
\newblock \bibinfo{journal}{\emph{The New York Times}} (\bibinfo{year}{2022}).
\newblock
\urldef\tempurl%
\url{https://www.nytimes.com/interactive/2022/05/13/us/covid-deaths-us-one-million.html}
\showURL{%
\tempurl}


\bibitem[White(1900)]%
        {white1900report}
\bibfield{author}{\bibinfo{person}{J~William White}.}
  \bibinfo{year}{1900}\natexlab{}.
\newblock \showarticletitle{Report of the Committee of the American Surgical
  Association on the Medicolegal Relations of the X-Rays}.
\newblock \bibinfo{journal}{\emph{Buffalo Medical Journal}}
  \bibinfo{volume}{40}, \bibinfo{number}{2} (\bibinfo{year}{1900}),
  \bibinfo{pages}{121}.
\newblock


\bibitem[Williamson et~al\mbox{.}(2009)]%
        {williamson2009body}
\bibfield{author}{\bibinfo{person}{Laila Williamson}, \bibinfo{person}{Serinity
  Young}, \bibinfo{person}{Janet Gyatso}, {and} \bibinfo{person}{Romio
  Shrestha}.} \bibinfo{year}{2009}\natexlab{}.
\newblock \bibinfo{booktitle}{\emph{Body \& Spirit: Tibetan Medical
  Paintings}}.
\newblock \bibinfo{publisher}{University of Washington Press},
  \bibinfo{address}{Seattle, WA}.
\newblock


\bibitem[Worlds(2023)]%
        {bodyworlds}
\bibfield{author}{\bibinfo{person}{Body Worlds}.}
  \bibinfo{year}{2023}\natexlab{}.
\newblock \bibinfo{title}{Fascination Body Worlds}.
\newblock \bibinfo{howpublished}{\url{https://bodyworlds.com/exhibitions/}}.
\newblock


\bibitem[Wysocki and Johnson-Eilola(1999)]%
        {wysocki_blinded_1999}
\bibfield{author}{\bibinfo{person}{Anne~Frances Wysocki} {and}
  \bibinfo{person}{Johndan Johnson-Eilola}.} \bibinfo{year}{1999}\natexlab{}.
\newblock \showarticletitle{Blinded by the {Letter}:}.
\newblock In \bibinfo{booktitle}{\emph{Passions {Pedagogies} and 21st {Century}
  {Technologies}}}, \bibfield{editor}{\bibinfo{person}{Gail~E. Hawisher} {and}
  \bibinfo{person}{Cynthia~L. Selfe}} (Eds.). \bibinfo{publisher}{University
  Press of Colorado}, \bibinfo{pages}{349--368}.
\newblock
\showISBNx{978-0-87421-258-7}
\urldef\tempurl%
\url{https://doi.org/10.2307/j.ctt46nrfk.22}
\showDOI{\tempurl}


\bibitem[Xiang and Venkatesan(2023)]%
        {xiang2023role}
\bibfield{author}{\bibinfo{person}{Jinpo Xiang} {and} \bibinfo{person}{Santhana
  Venkatesan}.} \bibinfo{year}{2023}\natexlab{}.
\newblock \showarticletitle{The role of Vesalius and his contemporaries in the
  transfiguration of human anatomical science}.
\newblock \bibinfo{journal}{\emph{Journal of Anatomy}} \bibinfo{volume}{242},
  \bibinfo{number}{2} (\bibinfo{year}{2023}), \bibinfo{pages}{124--131}.
\newblock


\bibitem[Zimmer(2005)]%
        {zimmer2005soul}
\bibfield{author}{\bibinfo{person}{Carl Zimmer}.}
  \bibinfo{year}{2005}\natexlab{}.
\newblock \bibinfo{booktitle}{\emph{Soul Made Flesh}}.
\newblock \bibinfo{publisher}{Simon and Schuster}.
\newblock


\end{thebibliography}

\appendix
\section*{Image Sourcing and Licensing}
\noindent
\textbf{\autoref{fig:fivefigures}}: Public domain, from \href{https://www.gutenberg.org/files/46572/46572-h/images/plate-xxxiii.jpg}{Project Gutenberg}, Studies in the History and Method of Science~\cite{gutenberg2014historymethods}.

\noindent
\textbf{\autoref{fig:mansur}}: Both images public domain, from Islamic Medical Manuscripts collection at the \href{https://www.nlm.nih.gov/hmd/arabic/p18.html}{U.S. National Library of Medicine} collection.

\noindent
\textbf{\autoref{fig:neijingtu}}: Public domain, from \href{https://commons.wikimedia.org/wiki/File:NeijingTu1.jpg}{Wikimedia Commons}. 

\noindent
\textbf{\autoref{fig:modernschema}, left panel}: From the authors, inspired after Netter Images' \href{https://www.netterimages.com/cardiovascular-system-overview-labeled-hansen-physiology-1e-physiology-james-a-perkins-4972.html}{Cardiovascular System Overview}. 

\noindent
\textbf{\autoref{fig:modernschema}, right panel}: Photograph by PaRappa 276, Creative Commons Attribution-Share Alike 4.0, from \href{https://commons.wikimedia.org/wiki/File:1960\%27s_edition_of_Operation.jpg}{Wikimedia Commons}.

\noindent
\textbf{\autoref{fig:easternanatomy}, top panel}: 
  Public domain, from the \href{https://www.nlm.nih.gov/exhibition/historicalanatomies/Images/1200_pixels/hua_t06.jpg}{U.S. National Library of Medicine} collection.

\noindent
\textbf{\autoref{fig:easternanatomy}, bottom panel}: 
Public domain, from \href{https://commons.wikimedia.org/wiki/File:The_Blue_Beryl-Anatomy_Vulnerable_Points.jpg}{Wikimedia Commons}.

\noindent
\textbf{\autoref{fig:musclemen}}: Public domain from the \href{https://www.nlm.nih.gov/exhibition/historicalanatomies/Images/1200_pixels/Vesalius_Pg_174.jpg}{U.S. National Library of Medicine} collection.

\noindent
\textbf{\autoref{fig:vr-ai-assistant}}: from work of Chheang et al.~\cite{chheang2023towards}, used with permission of the authors. 

\noindent
\textbf{\autoref{fig:humors}}:
Public domain, from \href{https://commons.wikimedia.org/wiki/File:Quinta_Essentia_(Thurneisse)_illustration_Alchemic_approach_to_four_humors_in_relation_to_the_four_elements_and_zodiacal_signs.jpg}{Wikimedia Commons}.

\noindent
\textbf{\autoref{fig:woundman}, top panel}: 
Creative Commons BY 4.0 International from the Wellcome Collection. \href{https://commons.wikimedia.org/wiki/File:Anathomia._53v_\%27Wound_man\%27;_flesh_tinted;_weapons_coloured._Wellcome_L0045155.jpg}{Wellcome MS290 L0045155}. 

\noindent
\textbf{\autoref{fig:woundman}, bottom panel}: 
Creative Commons BY 4.0 International from the Wellcome Collection.\href{https://commons.wikimedia.org/wiki/File:Anathomia._52v_Pregnant_woman,_seated_legs_apart_Wellcome_L0045195.jpg}{Wellcome MS290 L0045195}.

\noindent
\textbf{\autoref{fig:tibetandiagnostics}}:
Public domain, via \href{https://commons.wikimedia.org/wiki/File:The_Blue_Beryl-Methods_of_Treatment.jpg}{Wikimedia Commons}.

\noindent
\textbf{\autoref{fig:bruises}}:
Work by Giorgia Lupi and Kaki King, design credited to Giorgia Lupi and Pentagram, used with permission.

\noindent
\textbf{\autoref{fig:venuses}, left panel}: 
Public domain from the \href{https://www.nlm.nih.gov/exhibition/historicalanatomies/Images/1200_pixels/Gautier_angel.jpg}{U.S. National Library of Medicine} collection.

\noindent
\textbf{\autoref{fig:venuses}, right panel}: 
The Museum of Palazzo Poggi, specific image in this paper Creative Commons Attribution-Share Alike Generic 2.0 from \href{https://commons.wikimedia.org/wiki/File:Museo_Palazzo_Poggi.jpg}{Wikimedia Commons}.

\noindent
\textbf{\autoref{fig:albinus}}: Public domain from the \href{https://www.nlm.nih.gov/exhibition/historicalanatomies/Images/1200_pixels/Albinus_t04.jpg}{U.S. National Library of Medicine} collection.

\noindent
\textbf{\autoref{fig:grayropinski}, left panel}: 
Public domain, from \href{https://commons.wikimedia.org/wiki/File:Surgical_Anatomy_of_the_Arteries_of_the_neck_Gray\%27s_Anatomy_1858.jpg}{Wikimedia Commons}. 

\noindent
\textbf{\autoref{fig:grayropinski}, right panel}: 
From presentation by Ropinksi et al.~\cite{ropinski09illustravis}, used with permission of the authors.


\end{document}